\documentclass[12pt]{iopart}

\expandafter\let\csname equation*\endcsname\relax
\expandafter\let\csname endequation*\endcsname\relax
\usepackage{multirow}
\usepackage{glossaries-extra}
\usepackage{booktabs}
\usepackage{xcolor}
\usepackage{graphicx} % Required for inserting images
\usepackage{dcolumn}% Align table columns on the decimal point
\usepackage{bm}% bold math
\usepackage{physics}
\usepackage{amssymb}
\usepackage{mathtools}
\usepackage{hyperref}
\usepackage{subcaption}
\usepackage{pdflscape}
\usepackage[
    backend=biber,
    style=phys,
  ]{biblatex}
\newabbreviation{fp}{FP}{fixed point}
\newabbreviation{fpe}{FPE}{fixed-point equation}
\newabbreviation{qfp}{QFP}{quasi-fixed point}
\newabbreviation{nprg}{NPRG}{nonperturbative renormalization group}
\newabbreviation{frg}{fRG}{functional renormalization group}
\newabbreviation{ivp}{IVP}{initial value problem}
\newabbreviation{sm}{SM}{shooting method}
\newabbreviation{mol}{MOL}{method of lines}
\newabbreviation[% options to override defaults
  longplural={universality classes},
]{uc}{UC}{universality class}
\newabbreviation{rg}{RG}{renormalization group}
\newabbreviation{mc}{MC}{Monte Carlo}
\newabbreviation{qlro}{QLRO}{quasi-long-range ordered}
\newabbreviation{lro}{LRO}{long-range ordered}
\newabbreviation{pmc}{PMC}{principle of maximal conformity}
\newabbreviation{pms}{PMS}{principle of minimal sensitivity}
\newabbreviation{kt}{KT}{Kosterlitz-Thouless}
\newabbreviation{nf}{NF}{Nelson and Fisher}
\newabbreviation{bzj}{BZJ}{Br\'ezin and Zinn-Justin}
\newabbreviation{ch}{CH}{Cardy and Hamber}
\newabbreviation{de}{DE}{derivative expansion}
\newabbreviation{lpa}{LPA}{local potential approximation}
\newabbreviation{uv}{UV}{ultraviolet}
\newabbreviation{ir}{IR}{infrared}

\newabbreviation{wp}{WP}{Wilson-Polchinski}
\newabbreviation{mw}{MW}{Morris-Wetterich}

\newabbreviation{nr}{NR}{Newton-Raphson}
\newabbreviation{rk}{RK}{Runge-Kutta}

\newabbreviation{lce}{LCE}{large charge expansion}
\newabbreviation{pt}{PT}{perturbation theory}

\newcommand{\mycomment}[1]{}

\begin{document}

\title[Numerical Precision of the Derivative-Expansion-Based Functional RG]{Numerical Accuracy of the Derivative-Expansion-Based Functional Renormalization Group}
\author{Andrzej Chlebicki${ }^1$ }
\address{${ }^1$ Institute of Theoretical Physics, Faculty of Physics, University of Warsaw,\\
Pasteura 5, 02-093 Warsaw, Poland}
\ead{chlebickiandrzej@gmail.com}
\begin{abstract}
We investigate the precision of the numerical implementation of the functional renormalization group based on extracting the eigenvalues from the linearized RG transformation. For this purpose, we implement the LPA and $O(\partial^2)$ orders of the derivative expansion for the three-dimensional $O(N)$ models with $N~\in~\{1,2,3\}$. We identify several categories of numerical error and devise simple tests to track their magnitude as functions of numerical parameters. Our numerical schemes converge properly and are characterized by errors of several orders of magnitude smaller than the error bars of the derivative expansion for these models. We highlight situations in which our methods cease to converge, most often due to rounding errors. In particular, we observe an impaired convergence of the discretization scheme when the $\tilde \rho$ grid is cut off at the value $\tilde \rho_{\text{Max}}$ smaller than $3.5$ times the local potential minimum. The program performing the numerical calculations for this study is shared as an open-source library accessible for review and reuse.
\end{abstract}
\vspace{2pc}
\submitto{\JSTAT}
\maketitle
\section{Introduction}
Over the past 30 years, the \gls{frg} scheme based on the \gls{mw} equation has proven an efficient and versatile tool for investigating phenomena involving physics at various length scales. It was applied in a wide range of fields ranging from statistical and condensed-matter physics to gravity and high-energy physics. The success of this approach is connected to the development of well-controlled approximation schemes not relying on perturbative expansions, such as the \gls{de} \cite{Wetterich1993, Morris1994, Morris1997}. The nonperturbative nature of this scheme, however, comes at a significant cost. The \gls{rg} flow equations, obtained with this approximation, take the form of nonlinear partial differential equations that seldom can be solved analytically. Consequently, most recent studies employing the \gls{de} in the \gls{mw} approach rely on numerical calculations.

In recent years, significant attention has been directed towards forming a foundational understanding of the \gls{de}. The major developments include the first estimation of the convergence parameter \cite{Balog2019}, a method for calculating error bars \cite{DePolsi2020}, and a new rationale for choosing the infrared regulator based on the principle of maximal conformity \cite{Balog2020}. These developments are part of a concerted effort to elevate the status of the \gls{de} to a reliable tool delivering well-controlled qualitative predictions. In this spirit, we find it paramount to investigate how the numerical errors and the choice of numerical methods affect the results of \gls{de} calculations.

The \gls{frg} literature tends to omit numerical methods. Especially in older articles, there are few references to techniques or details of numerical implementation and numerical results are often treated as exact. Nowadays, the situation is improving and more information is being provided. Still, however, only a few papers present all the information necessary to reproduce the numerical results faithfully. The lack of transparency concerning numerical implementation might be hindering the development of our methodology.

Recently, numerical methods, in the \gls{frg} context, have seen more attention. For example, Refs.~\cite{Borchardt2016, Aoki2018, Grossi2021, Koenigstein2022, Beyer2022, Ihssen2023a, Ihssen2023b} implement new integration techniques for the \gls{frg} flows in various physical contexts. These works, however, focused exclusively on the \gls{rg}-flow integration. In the present work, we investigate a method of extracting critical exponents based on linearized \gls{frg} flow around a \gls{fp}, which is a subject of far fewer developments, e.g. \cite{Borchardt2015, Tan2022}. More specifically, we implement and evaluate the numerical errors of the stability-matrix approach based on the finite-grid representation of the functions parametrizing the effective average action and extracting the \gls{rg} eigenvalues from the spectrum of the operator describing the linearized flow.

The purpose of the present paper is threefold. Firstly, we aim to evaluate some of the numerical methods often employed in calculations based on the \gls{de} and present simple tests that may be used to examine their reliability. Secondly, we identify several sources of numerical error in calculations of this type and show how numerical parameters can be tuned to maximize precision. Finally, the program used to perform all the calculations for the purpose of this study is shared for review in the form of a simple \textit{Mathematica} \cite{Mathematica} library that can be relatively easily adapted for other purposes. Our analysis is conducted in the paradigmatic $O(N)$ models in three dimensions at the \gls{lpa} and $O(\partial^2)$ orders of the \gls{de}.

The structure of the paper is as follows. In Sec.~\ref{section:frg}, we introduce the \gls{mw} approach to the functional renormalization group and the approximation scheme of the \gls{de}. Sec.~\ref{section:numerical_methods} is devoted to the discussion of numerical techniques used in our calculations. In Sec.~\ref{section:results}, we present our results regarding the convergence rates and error propagation observed in our calculations. Finally, in Sec.~\ref{section:conclusion} we discuss our results and state our conclusions.

\section{Functional Renormalization Group}
\label{section:frg}
In the present chapter, we offer a brief overview of the \gls{mw} formulation of the \gls{frg} and the \gls{de}. For a more detailed description of this methodology and a review of its wide applications, see Refs.~\cite{Berges2002, Kopietz2010, Gies2012, Dupuis2020}.

The \gls{mw} approach to the \gls{frg} is centered around a scale-dependent functional of the order-parameter $\bm \phi$ called the effective average action $\Gamma_k$. The effective average action is defined in such a way that it smoothly interpolates between the microscopic action in the ultraviolet limit and the Gibbs free energy in the \gls{ir} limit. The flow of the effective average action is governed by the exact functional differential equation called the Morris-Wetterich equation
\begin{equation}
    k\partial_k \Gamma_k = \frac{1}{2} \rm{Tr} \left(k \partial_k R_k\right) \left(R_k + \Gamma^{(2)}\right)^{-1} \label{eq:morris_wetterich}.
\end{equation}
In the equation above, $R_k$ denotes the \gls{ir} regulator function, $\Gamma^{(2)}$ is the \gls{ir}-cutoff two-point function, and the trace is taken over all internal indices. 

In the \gls{frg} literature, we can find several commonly used \gls{ir} regulator functions $R_k$. In this work, we employ just two regulators:
\begin{itemize}
    \item the Litim regulator $R^{\text{Litim}}_k(\bm q^2) = Z^k k^2 \left(1-\frac{\bm q^2}{k^2}\right) \Theta \left(1-\frac{\bm q^2}{k^2}\right)$,
    \item the exponential regulator $R^{\text{Exp}}_k(\bm q^2) = \alpha Z^k k^2\exp\left(-\frac{\bm q^2}{k^2}\right)$.
\end{itemize}
In the definitions above, $\Theta$ stands for the Heaviside step function, $Z^k$ denotes the order-parameter-renormalization factor [to be defined shortly], and $\alpha$ is an arbitrary positive constant. We employ the Litim regulator at the \gls{lpa} level whenever we want to perform the integrals analytically and the exponential regulator otherwise. In calculations of physical quantities, the regulator constant $\alpha$ should be optimized via the \textit{principle of minimal sensitivity} or the \textit{principle of maximal conformity} to reduce the regulator dependence \cite{Canet2003, Balog2020}. In this work, however, we are only interested in the numerical error which can be assumed to be mostly independent of the value of $\alpha$. For this reason, we refrain from optimizing the regulator and assume the value $\alpha=1$.

The \gls{mw} equation \eqref{eq:morris_wetterich} is not directly soluble. The derivative expansion is an approximate scheme that recasts the \gls{mw} equation into a set of partial differential equations that can be solved using numerical methods. This approximation is based on an observation that, from the perspective of long-range physics, we are mostly interested in the low-momentum region of the correlation function. We, therefore, expand the effective average action in powers of the order-paramater's momentum around a spatially homogeneous order-parameter configuration.

In practice, the \gls{de} is set up by imposing an ansatz on the effective average action~$\Gamma_k$. A proper \gls{de} ansatz at the order $O(\partial^p)$ contains all terms respecting the required symmetries of up to $p$-th power of the spatial-derivative operator. In~the present work, we consider the $O(N)$-symmetric theories of an $N$-component scalar field. The leading- and next-to-leading-order \gls{de} ans\"atze for these models read:
\begin{subequations}
\label{eq:on_de_ansatze}
\begin{align}
    &\Gamma_k^{\mathrm{LPA}} = \int d^dx \left\{U^k(\rho) + \frac{1}{2} \left(\partial_\mu \phi^i\right)^2\right\}, \label{eq:on_lpa_ansatz}\\
    &\Gamma_k^{\partial^2} = \int d^dx \left\{U^k(\rho) + \frac{Z^k_\pi(\rho)}{2} \left(\partial_\mu \phi^i\right)^2 + \frac{Z^k_\sigma(\rho)-Z^k_\pi(\rho)}{4\rho} \left(\partial_\mu \rho\right)^2\right\}, \label{eq:on_de2_ansatz}
\end{align}
\end{subequations}
where the parametrizing functions $U^k$, $Z_\sigma^k$, and $Z_\pi^k$ depend on the order parameter $\bm \phi$ only through the $O(N)$ invariant $\rho = \frac{\phi^i \phi^i}{2}$. In such a parametrization, all the symmetry constraints are fulfilled by construction. To preserve the analyticity of the action, however, we have to impose an additional constraint $Z^k_\sigma(0)=Z^k_\pi(0)$. At the leading order, called the \gls{lpa}, the effective action is parametrized by just one function $U^k(\rho)$ - the local potential. The order $O(\partial^2)$ introduces two kinetic coefficients $Z^k_\sigma(\rho)$ and $Z^k_\pi(\rho)$%\footnote{In the Ising universality class [$N=1$], $Z^k_\pi(\rho)$ loses its physical significance. Consequently, there remain just two physically relevant independent parametrizing functions $U^k(\rho)$, $Z^k_\sigma(\rho)$. In the calculations of the present work, we perform the calculations for the $O(1)$ model including the nonphysical operators connected to $Z^k_\pi(\rho)$. This choice does not affect the physical eigenvalues except for numerical errors.}. 
\footnote{In the Ising universality class [$N=1$], the kinetic terms $\left(\partial_\mu \phi^1\right)^2$ and $\left(\partial_\mu \rho\right)^2 = 2 \rho \left(\partial_\mu \phi^1\right)^2$ are proportional. The ansatz \eqref{eq:on_de2_ansatz} is constructed in such a way that, for $N=1$, $Z^k_\pi(\rho)$ drops out of the ansatz and loses its physical significance. Consequently, there remain just two physically relevant independent parametrizing functions $U^k(\rho)$, $Z^k_\sigma(\rho)$. In the calculations of the present work, we perform the calculations for the $O(1)$ model including the nonphysical operators connected to $Z^k_\pi(\rho)$. This choice does not affect the physical eigenvalues except for numerical errors.}. 
Many studies also employ the so-called \gls{lpa}' truncation, which is an intermediate step between \gls{lpa} and $O(\partial^2)$ orders of the \gls{de}. In this truncation, the kinetic terms are reduced to flowing constants $Z_\sigma^k(\rho) = Z_\sigma^k$, $Z_\pi^k(\rho) = Z_\pi^k$, or to a single flowing constant $Z_\sigma^k = Z_\pi^k=Z^k$. The \gls{de} was successfully applied to the $O(N)$ models up to the order $O(\partial^4)$ for generic values of $N$ and up to the order $O(\partial^6)$ for $N=1$\cite{Balog2019, DePolsi2020}.

We derive the flow equations by plugging the effective action ans\"atze into the \gls{mw} equation \eqref{eq:morris_wetterich}. The flow of the local potential is derived by evaluating Eq.~\eqref{eq:morris_wetterich} in a spatially uniform field configuration 
\begin{equation}
    \phi^i_{\bm q}= \phi \delta_{i,1} \delta_{\bm q, 0},
\end{equation}
while the flow of kinetic coefficients is extracted from the flow of $\Gamma^{(2)}$ vertices via:
\begin{subequations}
\label{eq:de2_dimensionful}
    \begin{align}
        &\beta^k_U(\rho) = k \partial_k U^k(\rho) = \left. k \partial_k \Gamma_k[\bm \phi]\right|_{\text{Uniform}}, \\
        &\beta^k_{Z_\sigma}(\rho) = k \partial_k Z^k_\sigma(\rho) = \frac{1}{2d} \Delta_{\bm p}\left.  \left(k \partial_k \Gamma^{(2)}_{k;1 \bm{p},1 -\bm{p}}\right)\right\vert_{\text{Uniform, } \bm p=0}, \\
        &\beta^k_{Z_\pi}(\rho) = k \partial_k Z^k_\pi(\rho) = \frac{1}{2d} \Delta_{\bm p} \left.  \left(k \partial_k \Gamma^{(2)}_{k;2 \bm{p},2 -\bm{p}}\right)\right\vert_{\text{Uniform, } \bm p=0}.
    \end{align}
\end{subequations}
The $\beta$ functions defined in Eq.~\eqref{eq:de2_dimensionful} do not depend on the local potential directly, only through its derivatives. It is, therefore, convenient to reframe the problem in terms of $V^k(\rho) = \partial_\rho U^k(\rho)$. The $\beta$ function for the derivative of the local potential reads:
\begin{align}
    \beta^k_{V}(\rho) = -\frac{1}{2} \int \frac{d^d\bm q}{\left(2\pi\right)^d} \left[k\partial_k R_k\left(\bm q^2\right)\right] &\Bigg\{\frac{3 V'(\rho ) + 2 \rho  V''(\rho )+\bm q^2 Z_\sigma'(\rho )}{\left[V(\rho) + 2 \rho V'(\rho) + Z_\sigma(\rho) \bm q^2 + R_k(\bm q^2)\right]^2} \nonumber \\
    &+ (N-1)\frac{V'(\rho ) + \bm q^2 Z_\pi'(\rho )}{\left[V(\rho) + Z_\pi(\rho) \bm q^2 + R_k(\bm q^2)\right]^2}\Bigg\}, \label{eq:v_flow}
\end{align}
where the scale $k$ indices were suppressed for clarity. We refrain from presenting the $\beta$ functions for the kinetic coefficients, as they are quite lengthy. They are presented in the online repository serving as an appendix to this article \cite{Chlebicki2024Repository}.

In the present work, we investigate the linearized \gls{rg} flow around a \gls{fp} solution. To make the fixed-point behavior manifest we express the equations in terms of dimensionless variables by performing the following rescaling:
\begin{alignat}{3}
    &\tilde \rho = \rho k^{2-d} Z^k, \quad &&\tilde V^t(\tilde \rho) = V^k(\rho) k^{-2}  {Z^k}^{-1}, \quad &&\tilde{Z}^t_{\sigma/\pi}(\tilde \rho) = Z^k_{\sigma/\pi}(\rho) {Z^k}^{-1}, \label{eq:rescaling}\\
    &t = -\log\left(k/\Lambda\right)\footnotemark, \quad && \tilde \beta_{V}(\tilde \rho) = \beta^k_{V}(\rho) k^{-2}  {Z^k}^{-1}, \quad && \tilde{\beta}_{Z_{\sigma/\pi}}(\tilde \rho) = \beta^k_{Z_{\sigma/\pi}}(\rho)  {Z^k}^{-1}. \nonumber
\end{alignat}
\footnotetext{We note, that in the present convention the \textit{renormalization time} $t$ flows in the positive direction from $0$ to $+\infty$, with the scale $k$ decreasing from $\Lambda$ to $0$.} The order-parameter-renormalization factor $Z^k$, used in the rescaling, is defined by normalizing one of the dimensionless kinetic factors [typically $\tilde Z_\pi$] to unity at an arbitrary point $\tilde \rho_\eta$
\begin{equation}
    \tilde Z^t_{\pi}(\tilde \rho_\eta) = Z^k_{\pi}(\rho_\eta)/Z^k \equiv 1.
\end{equation}
In the present work, we choose $\rho_\eta = 0$, which conveniently removes the choice between normalizing $\tilde Z_\sigma$ and $\tilde Z_\pi$ since $Z^k_\sigma(0) = Z^k_\pi(0)$. We also define the \textit{running anomalous dimension} $\eta_t$, serving as a precursor to the fixed-point anomalous dimension $\eta^*$, as the logarithmic derivative of the order-parameter-renormalization factor $\eta_t = \partial_t \log(Z^k)$.

\mycomment{
In the present work, we investigate the linearized \gls{rg} flow around a \gls{fp} solution. To make the fixed-point behavior manifest we express the equations in terms of dimensionless variables by performing the following rescaling:
\begin{alignat}{3}
    &\tilde \rho = \rho k^{2-d -\eta_k}, \quad &&\tilde V^t(\tilde \rho) = V^k(\rho) k^{2-\eta_k}, \quad &&\tilde{Z}^t_{\sigma/\pi}(\tilde \rho) = Z^k_{\sigma/\pi}(\rho) k^{\eta_k}, \label{eq:rescaling}\\
    &t = -\log\left(k/\Lambda\right), \quad && \tilde \beta_{V}(\tilde \rho) = \beta^k_{V}(\rho) k^{2-\eta_k}, \quad && \tilde{\beta}_{Z_{\sigma/\pi}}(\tilde \rho) = \beta^k_{Z_{\sigma/\pi}}(\rho) k^{\eta_k}. \nonumber
\end{alignat}
The running anomalous dimension $\eta_k$ used in the rescaling is defined as the logarithmic derivative of the order-parameter-renormalization factor $\eta_k = \partial_t \log(Z^k)$. The set of equations is closed by defining $Z^k$ by normalizing one of the dimensionless kinetic factors [typically $\tilde Z_\pi$] to unity at an arbitrary point $\tilde \rho_\eta$
\begin{equation}
    \tilde Z^t_{\pi}(\tilde \rho_\eta) = Z^k_{\pi}(\rho_\eta)/Z^k \equiv 1.
\end{equation}
In the present work, we choose $\rho_\eta = 0$, which conveniently removes the choice between normalizing $\tilde Z_\sigma$ and $\tilde Z_\pi$ since $Z^k_\sigma(0) = Z^k_\pi(0)$.
}

In terms of the dimensionless variables, the flow equations take the form:
\begin{subequations}
\label{eq:dimensionless_form}
\begin{alignat}{4}
    &\partial_t \tilde V( \tilde \rho) &&= (\eta_t-2) \tilde V(\tilde \rho) &&- (d-2+\eta_t) \tilde \rho \tilde V'(\tilde \rho) &&- \tilde \beta_V(\tilde \rho), \\
    &\partial_t \tilde Z_\sigma( \tilde \rho) &&= \eta_t \tilde Z_\sigma(\tilde \rho) &&- (d-2+\eta_t) \tilde \rho \tilde Z_\sigma'(\tilde \rho) &&- \tilde \beta_{Z_\sigma}(\tilde \rho),\\
    &\partial_t \tilde Z_\pi( \tilde \rho) &&= \eta_t \tilde Z_\pi(\tilde \rho) &&- (d-2+\eta_t) \tilde \rho \tilde Z_\pi'(\tilde \rho) &&- \tilde \beta_{Z_\pi}(\tilde \rho),
\end{alignat}
\end{subequations}
in which the explicit renormalization-time $t$ dependence has been removed. In the form~\eqref{eq:dimensionless_form}, the flow equations can be used to find the \gls{fp} solutions
\begin{equation}
    \partial_t \mathcal{F}^* = \partial_t \begin{pmatrix}
        \tilde V^* \\ \tilde Z_\sigma^* \\\tilde Z_\pi^*
    \end{pmatrix} = 0. \label{eq:fp_equation}
\end{equation}

Once the \gls{fp} solution $\mathcal{F}^*$ has been identified, the flow equations can be linearized to determine the \gls{rg} eigenvalues which can be used to calculate the critical exponents. The linearization is performed by perturbing the \gls{fp} solution:
\begin{equation}
    \mathcal{F}_\epsilon = \mathcal{F}^* + \epsilon \begin{pmatrix}
        f_{V} \\ f_{Z_\sigma} \\ f_{Z_\pi}
    \end{pmatrix}, \label{eq:fp_perturbation}
\end{equation}
with $f_{Z_\pi}(t,\tilde \rho_\eta) = 0$ and $f_{Z_\pi}(t,\tilde \rho= 0) = f_{Z_\sigma}(t,\tilde \rho= 0)$. The perturbed \gls{fp} action \eqref{eq:fp_perturbation} is inserted into the flow equations \eqref{eq:dimensionless_form}, which are subsequently truncated at the linear order in $\epsilon$. After simplification, the resulting equations take the linear form:
\begin{equation}
    \partial_t \begin{pmatrix}
        f_V(t,\tilde \rho) \\ f_{Z_\sigma}(t,\tilde \rho) \\ f_{Z_\pi}(t,\tilde \rho)
    \end{pmatrix} = \mathcal{M}(\tilde \rho) \begin{pmatrix}
        f_V(t,\tilde \rho) \\ f_{Z_\sigma}(t,\tilde \rho) \\ f_{Z_\pi}(t,\tilde \rho)
    \end{pmatrix}, \label{eq:rg_linearized}
\end{equation}
where $\mathcal{M}$ is a differential operator to which we shall refer as the stability operator [or stability matrix in a discretized form].

The stability operator $\mathcal{M}$ is an unbounded and asymmetrical differential operator. To our knowledge, $\mathcal{M}$ has not been comprehensively investigated from the perspective of functional analysis. Despite its incomplete mathematical understanding, the stability operator has proven to be an effective tool for calculating critical exponents. Its eigenvalues $\{e_i\}$ represent the scaling dimensions of their associated eigenoperators [or eigenvectors] $F_i$. When the investigated \gls{fp} is critical, the leading and only positive eigenvalue of $\mathcal{M}$ is an inverse of the correlation length exponent $e_1 = \nu^{-1}$, while the remaining eigenvalues, in principle, determine the correction-to-scaling exponents $\omega_{i} = \abs{e_{i+1}}$. 

It should be noted that not all of the eigenvalues $e_i$ affect the scaling behavior of physical observables. Some of the eigenvectors $F_i$, known as the redundant operators, are connected to a reparametrization of the effective action $\Gamma$ and are therefore an artifact of the employed scheme. This observation forms a basis of the so-called essential scheme of the \gls{de} \cite{Baldazzi2022}. In this approach, the redundant operators associated with nonlinear transformations of the order parameter are removed from the parametrization of the effective action. As a result, the spectrum of the linear \gls{rg} operator calculated in the essential scheme does not include these redundant operators. From the perspective of numerical accuracy, however, the distinction between essential and redundant operators does not seem meaningful. We, therefore, implement the simpler ``standard'' scheme and do not discuss whether the calculated eigenvalues are redundant.

\section{Numerical Methods}
\label{section:numerical_methods}
The \gls{frg} \glspl{fpe} form a set of ordinary differential equations. In the literature, we can find several different approaches to solving them numerically. One of the techniques found in the literature is the so-called \gls{sm}, in which the \glspl{fpe} are integrated as an initial value problem. 

The \gls{sm} aims to answer the question of how to specify the initial condition corresponding to a specific global fixed-point solution. Locally, almost any initial condition leads to a local solution of the \glspl{fpe}. However, only a discrete number of initial conditions lead to physically relevant global solutions \cite{Morris1995, Codello2012}. The \gls{sm} is an iterative scheme in which the initial value problem is solved for many different initial conditions. With each iteration, the initial condition is systematically improved until the obtained solution of the \glspl{fpe} specific lies closer to the global \gls{fp} solution than a given tolerance threshold. After identifying the \gls{fp} solution, the \gls{rg} eigenvalues can be identified using the \gls{sm} to solve the eigenvalue equation:
\begin{equation}
    \mathcal{M}(\tilde \rho) F_i(\tilde \rho) = \lambda_i F_i(\tilde \rho)
\end{equation}
for the stability operator $\mathcal{M}$ [see Eq.~\eqref{eq:rg_linearized}]. For more details on how this method can be implemented, see Refs. \cite{Codello2012, Tan2022}.

The \gls{sm} was used for studying the critical and multicritical behavior of the $O(N)$ models. In the Ising universality class [$N=1$] this method was applied at the \gls{lpa} order, where the initial condition is specified by a single variable, and at the $O(\partial^2)$ order, where the initial condition consists of two variables \cite{Morris1995, Bervillier2008, Defenu2018}. Similarly, for the generic value of $N$ the \gls{sm} was used at the \gls{lpa} level [single-variable initial condition] and at the \gls{lpa}' level [two-variable initial condition] \cite{Codello2013, Codello2015, Yabunaka2017, Yabunaka2018, Tan2022}. Curiously, we are unaware of any study implementing the \gls{sm} at higher orders. This is probably due to the rapidly rising number of dimensions in the space of initial conditions. We finally note that although the intuitive approach is to integrate the \glspl{fpe} from $\tilde \rho=0$ towards $\tilde \rho \to \infty$ it is not the only possible solution. Ref.~\cite{Tan2022} identifies the asymptotic boundary condition at $\tilde \rho \to \infty$ and implements the \gls{sm} integrating the \glspl{fpe} from some large value $\tilde \rho_{\text{Max}}$ towards $\tilde \rho=0$.

\subsection{Finite-grid representation}
An alternative approach to solving the \glspl{fpe} relies on representing the functions parametrizing the effective average action in a finite-dimensional space. Such a discretization recasts the ordinary differential equations into a large set of algebraic equations that can be solved with various methods e.g. the Newton-Raphson method. There are many ways in which such a finite-dimensional representation can be implemented, see e.g. Refs.~\cite{Borchardt2015, Borchardt2016} employing Chebyshev polynomials for that purpose. In this work, however, we will study a more common and straightforward approach based on the grid representation.

Let us consider a $N_{\rho}+1$-point uniformly-spaced grid $\{\tilde \rho_i = i h_\rho\}_{i \in \overline{0, N_{\rho}}}$ spanning an interval $[0, \tilde \rho_{\text{Max}}]$ with $\tilde \rho_{\text{Max}} = h_\rho N_{\rho}$. Subsequently, let us reduce the domain of each parametrizing function $f$ to the discrete set of grid points $\{\tilde \rho_i = i h_\rho\}$, thus reducing $f$ to a vector of values $\left(f(\tilde\rho_0), \ldots, \tilde f(\tilde \rho_{N_{\rho}})\right)$. In a grid representation, the $\tilde \rho$~derivatives cannot be calculated analytically and have to be approximated with finite differences.

There are several different ways of constructing finite-difference approximations for the derivatives. In this work, we adopt a simple scheme, in which the derivative at the point $\tilde \rho_i$ is approximated by a linear combination of function values at $n$ grid points closest to $\tilde \rho_i$ for $n$ equal $3$, $5$, $7$, or $9$. The coefficients defining a linear combination approximating the $d$-th derivative $f^{(d)}(\tilde \rho_i)$ can be derived directly from the Taylor expansion, see Ref.~\cite{Fornberg1988}. For $n>d$, there is a unique set of coefficients defining an approximation of $f^{(d)}(\tilde \rho_i)$ accurate up to corrections of order $O\left(h_\rho^{n-d}\right)$\footnote{In some cases, when a central derivative formula is employed the approximation can be of the order $O\left(h_\rho^{n-d+1}\right)$ due to symmetry.}. In this scheme, the discrete derivative operators $D^d_n$ can be represented with matrices acting on the vectors of function values. E.g. for $N_\rho=4$, the $3$-point discrete derivative operators take the form:
\begin{subequations}
\begin{align}
&D_3^1 f = \frac{1}{h_\rho}\left(
\begin{array}{ccccc}
 -\frac{3}{2} & 2 & -\frac{1}{2} & 0 & 0 \\
 -\frac{1}{2} & 0 & \frac{1}{2} & 0 & 0 \\
 0 & -\frac{1}{2} & 0 & \frac{1}{2} & 0 \\
 0 & 0 & -\frac{1}{2} & 0 & \frac{1}{2} \\
 0 & 0 & \frac{1}{2} & -2 & \frac{3}{2} \\
\end{array}
\right)
\begin{pmatrix}
    f(\tilde \rho_0) \\ f(\tilde \rho_1) \\ f(\tilde \rho_2) \\ f(\tilde \rho_3) \\ f(\tilde \rho_4)
\end{pmatrix} = \begin{pmatrix}
    f^{(1)}(\tilde \rho_0) \\ f^{(1)}(\tilde \rho_1) \\f^{(1)}(\tilde \rho_2) \\ f^{(1)}(\tilde \rho_3) \\f^{(1)}(\tilde \rho_4)
\end{pmatrix} + O(h_\rho^2), \\
&D_3^2f = \frac{1}{h_\rho^2}\left(
\begin{array}{ccccc}
 1 & -2 & 1 & 0 & 0 \\
 1 & -2 & 1 & 0 & 0 \\
 0 & 1 & -2 & 1 & 0 \\
 0 & 0 & 1 & -2 & 1 \\
 0 & 0 & 1 & -2 & 1 \\
\end{array}
\right)
\begin{pmatrix}
    f(\tilde \rho_0) \\ f(\tilde \rho_1) \\ f(\tilde \rho_2) \\ f(\tilde \rho_3) \\ f(\tilde \rho_4)
\end{pmatrix} = \begin{pmatrix}
    f^{(2)}(\tilde \rho_0) \\ f^{(2)}(\tilde \rho_1) \\f^{(2)}(\tilde \rho_2) \\ f^{(2)}(\tilde \rho_3) \\f^{(2)}(\tilde \rho_4)
\end{pmatrix} + \begin{pmatrix}
    O(h_\rho) \\ O(h_\rho^2) \\ O(h_\rho^2) \\ O(h_\rho^2) \\ O(h_\rho)
\end{pmatrix}.
\end{align}
\label{eq:finite_difference_operators}
\end{subequations}
Higher-order discrete derivative operators employed in our calculations are presented in \ref{appendix:discrete_derivatives}. Note that the corrections to the second-derivative approximation are of order $O(h_\rho^2)$ in bulk, but only $O(h_\rho)$ at the borders. Higher accuracy of the bulk approximation arises due to the symmetry of the expressions which cannot be achieved at the border. Interestingly, we will see that the convergence rate for the estimates of the critical exponents is determined by the order of the corrections for the bulk approximation rather than the border expressions.

In the finite-grid representation, the \glspl{fpe} form a set of algebraic equations treating each grid-point value of each parametrizing function as an independent variable. This system of equations can be relatively easily solved with standard numerical methods; the Newton-Raphson method is a particularly efficient tool for this purpose. We note that the Newton-Raphson method requires an initial guess for the \gls{fp} solution which is improved in an iterative procedure. We have found, that to find the critical \gls{fp} in the three-dimensional $O(N)$ models it is often sufficient to adopt the initial guess with the ``Mexican-hat'' type potential and constant kinetic terms:
\begin{equation}
    \tilde U(\tilde \rho) = \frac{\tilde u}{2} (\tilde \rho - \tilde \rho_0)^2, \; \tilde Z_\sigma(\tilde \rho)=\tilde Z_\pi(\tilde\rho)=1,
\end{equation}
for some constants $\tilde u$ and $\tilde \rho_0$. We also note that this scheme is easily generalized to more complex settings. It has been recently implemented both at higher orders of the \gls{de} \cite{Balog2019, DePolsi2020}, as well as, for calculations involving more than one invariant \cite{Tissier2012, Delamotte2016b, Chlebicki2022}. 

\subsection{Method of lines}
\label{subsection:method_of_lines}
Let us recall Eq.~\eqref{eq:rg_linearized} describing the flow of an infinitesimal perturbation $\{\epsilon f_V(t,\tilde \rho), \epsilon f_{Z_\sigma}(t,\tilde \rho), \epsilon f_{Z_\pi}(t,\tilde \rho)\}$ around the fixed point $\mathcal{F}^*$:
\begin{equation}
    \partial_t \begin{pmatrix}
        f_V(t,\tilde \rho) \\f_{Z_\sigma}(t,\tilde \rho) \\ f_{Z_\pi}(t,\tilde \rho)
    \end{pmatrix} = \mathcal{M}(\tilde \rho) \begin{pmatrix}
        f_V(t,\tilde \rho) \\f_{Z_\sigma}(t,\tilde \rho) \\ f_{Z_\pi}(t,\tilde \rho)
    \end{pmatrix} \label{eq:rg_linearized2}
\end{equation}
with $\mathcal{M}(\tilde \rho)$ denoting a second-order differential operator called the stability operator. Linear partial differential equations such as Eq.~\eqref{eq:rg_linearized2} can be solved via the so-called \gls{mol}. In this approach, the $\tilde \rho$ space is discretized and, consequently, the stability operator~$\mathcal{M}(\tilde \rho)$ is approximated by a finite-dimensional stability matrix~$M$. Subsequently, the problem is either integrated numerically as an initial value problem or the spectrum $\{\lambda_i, F_i\}$ of $M$ is resolved to calculate time dependence analytically. 

In the \gls{rg} context, we are seldom interested in actual solutions to Eq.~\eqref{eq:rg_linearized2}. Oftentimes, we only want to obtain the eigenvalues of the stability operator $\mathcal{M}$ which are connected to the critical exponents. Nevertheless, we can draw important insights from the \gls{mol}. In particular, we note the well-known issue with the \gls{mol} that while several leading eigenvalues of $M$ offer a good approximation of the leading eigenvalues of $\mathcal{M}$ the accuracy quickly degrades for subsequent eigenvalues \cite{Suli2003, Butcher2016}. As discussed in Sec.~\ref{section:results}, our calculations confirm this prediction. For a detailed review of the \gls{mol}, see Ref.~\cite{Schiesser2009}. 

Since, in our approach, the \gls{fp} solution is already calculated in a discretized form, it is natural to use the same grid representation for the stability matrix. Let $\mathcal{F}_{a\tilde \rho_i}$ denote the value of the parametrizing function $a$ at the grid point $\tilde \rho_i$. The entries of the stability matrix are defined as:
\begin{equation}
    M_{a\tilde \rho_i,b\tilde \rho_j} \coloneq \left.\frac{\partial \left( \partial_t \mathcal{F}_{a\tilde \rho_i}\right)}{\partial \mathcal{F}_{b\tilde \rho_j} }\right|_{\mathcal{F} = \mathcal{F^*}}. \label{eq:stability_matrix}
\end{equation}
Calculating the stability matrix one has to mind that the anomalous dimension is not an independent variable and the terms of the form
\begin{equation}
    \left.\frac{\partial \left( \partial_t \mathcal{F}_{a\tilde \rho_i}\right)}{\partial \eta} \frac{\partial \eta }{\partial \mathcal{F}_{b\tilde \rho_j} }\right|_{\mathcal{F} = \mathcal{F^*}}
\end{equation}
also contribute to $M_{a\tilde \rho_i,b\tilde \rho_j}$. We also note that the value $\tilde Z_\pi(\tilde \rho_\eta)=1$ is fixed and, consequently, the row $M_{\tilde Z_\pi \tilde \rho_\eta, b\tilde \rho_j}$ and the column $M_{b\tilde \rho_j,\tilde Z_\pi \tilde \rho_\eta}$ should be removed from the matrix\footnote{There are implementations of this scheme in which one does consider the perturbation of $\tilde Z_\pi(\tilde \rho_\eta)$. This, however, introduces a redundant operator connected to an approximately zero eigenvalue, thus drastically increasing the condition number of $M$ and reducing the numerical accuracy when resolving its spectrum.}. Finally, analyticity of the effective action requires $\tilde Z_\sigma(0)=\tilde Z_\pi(0)$ and therefore the values $\tilde Z_\sigma(0)$ and $\tilde Z_\pi(0)$ should be treated as a single parameter.

%In the numerical framework, calculating the derivatives in Eq.~\eqref{eq:stability_matrix} analytically can be a difficult task and it is often convenient to approximate them with finite differences. 
The derivatives in Eq.~\eqref{eq:stability_matrix} can be calculated either analytically or numerically. Calculating the derivatives analytically can be very effective when implementing the \gls{de} in a symbolic environment like \textit{Mathematica}. This approach avoids approximation errors and can sometimes be more computationally efficient than the finite-difference approach. Beyond the order $O(\partial^2)$, however, the \glspl{fpe} become incredibly complex and calculating the stability matrix in a symbolic framework becomes essentially infeasible. In a non-symbolic setting [e.g. in \textit{C++} or \textit{Fortran}], calculating the derivatives analytically is a tedious task that, to our knowledge, has never been performed. In this case, it is common and significantly more convenient to approximate the derivatives with finite differences.

Like in the case of $\tilde \rho$~derivatives, the derivation of the finite-difference approximation is based on the Taylor expansion. The three lowest-order central approximations for the first derivative read:
\begin{align}
    f'(x) &= \frac{3}{4\epsilon }\left(f(x+\epsilon) - f(x-\epsilon)\right) - \frac{3}{20\epsilon }\left(f(x+2\epsilon) - f(x-2\epsilon)\right) \nonumber \\
    &+\frac{1}{60\epsilon }\left(f(x+3\epsilon) - f(x-3\epsilon)\right) + O(\epsilon^6) \nonumber \\
    &= \frac{2}{3\epsilon }\left(f(x+\epsilon) - f(x-\epsilon)\right) - \frac{1}{12\epsilon }\left(f(x+2\epsilon) - f(x-2\epsilon)\right) + O(\epsilon^4) \nonumber \\
    &= \frac{1}{2\epsilon }\left(f(x+\epsilon) - f(x-\epsilon)\right) + O(\epsilon^2). \label{eq:finite_difference_approximation}
\end{align}
We calculate the column $M_{ \cdot\,\cdot,b\tilde \rho_j}$ of the stability matrix by perturbing the function $b$ of the fixed point action at the point $\tilde \rho_j$ by a multiple of a small parameter $\epsilon$. Subsequently, we calculate the time derivative of each perturbed action and combine them using one of the formulas \eqref{eq:finite_difference_approximation} to obtain the column of the stability matrix. It is important to note that unlike in the $\tilde \rho$ space in this functional space, we do not introduce any grid representation. As a consequence, there are no boundary expressions like in Eq.~\eqref{eq:finite_difference_operators}, and the small parameter $\epsilon$ can be freely tuned to maximize the numerical precision.

In our calculations, we compute the derivatives defining the stability matrix analytically. The only exception is presented in Figs.~\ref{fig:stability_matrix_lpa} and \ref{fig:stability_matrix_de2} where we investigate the precision of the finite-difference approximation for the stability matrix.

\subsection{Loop integrals}
The loop integrals appearing in the \gls{frg} $\beta$ functions [see Eq.~\eqref{eq:v_flow}] can be performed analytically only for a specially selected \gls{ir} regulators. Unless such a regulator is employed, the integrals have to be approximated numerically. In the \gls{frg} literature, there are several methods commonly utilized for performing these integrals including the Simpson's rule and various Gaussian quadratures. In the present work, we perform integrals analytically at the \gls{lpa} level when the Litim regulator is used and employ the Gauss-Legendre quadrature otherwise. Unless specified otherwise, we perform the Gauss-Legendre integral on the interval $\abs{\bm{q}} \in [0, 5]$ with $35$ evaluation points ensuring that the integration error does not exceed $10^{-10}$ in any $\beta$ function at any grid point. For a detailed description of Gaussian quadrature rules, see to Ref.~\cite{Suli2003}.

For the investigated models, the integrals are single-loop $d$-dimensional spherically symmetric integrals. It is typical, to integrate out the spherical component analytically and reduce the integral to a one-dimensional integral with respect to $q=\abs{\bm {q}}$ or $y=\bm{q}^2$\footnote{In the $\beta$ functions in Ref.~\cite{Chlebicki2024Repository}, we include a constant $v_d=\frac{1}{2^{d+1} \pi^d \Gamma\left( d/2 \right)}$ which absorbs the angular-integration constant as well as the factor $\frac{1}{\left(2\pi\right)^d}$ [see Eq.~\eqref{eq:v_flow}]. In the calculations, however, we set $v_d=1$; this can be understood as the rescaling $\tilde \rho \to \tilde \rho/v_d$, $\Gamma \to \Gamma/v_d$, which does not affect the physical quantities. To avoid confusion regarding the rescaling convention we always measure $\tilde \rho_{\text{Max}}$ and $h_\rho$ in the units of the minimum of the fixed-point local potential $\tilde \rho_0$.}. In the case of three-dimensional models, the latter choice introduces a square-root-like nonanalyticity to the integrand around $y=0$. This is inconvenient from the perspective of numerical integration since most numerical integration techniques offer error bounds involving derivatives of the integrand, e.g. for the Simpson's 3/8 rule the error bound is proportional to the fourth derivative of the integrand at some point within the integration domain. Due to a nonanalyticity introduced by the choice of $y$ as the integration variable, the error bounds become very large and the convergence of the integral approximation becomes very slow. For this reason, we choose to integrate with respect to $q=\abs{\bm {q}}$. Although it is not directly relevant to this study, we note that in non-integer dimensions, the issue of nonanalyticity cannot be easily circumvented by a better choice of integration variable and has to be dealt with in some other way.

\subsection{Floating-point arithmetic}
In numerical calculations, real numbers are typically stored as binary numbers with a fixed finite number of digits. As a consequence, most real numbers cannot be represented exactly and are rounded when represented as a floating-point number thus introducing rounding errors. As we show in Sec.~\ref{section:results}, rounding errors are a relevant concern for high-precision calculations. Although in our calculations we use the \textit{Mathematica} software which allows for storing real numbers with arbitrary precision, we perform most of our calculations in a ``64-bit double'' format [represented by ``MachinePrecision'' in \textit{Mathematica}] offering a precision of approximately 16 decimal digits. This way we simulate the results that could be obtained if the same procedures were programmed in a low-level language like \textit{C++} or \textit{Fortran}. In some cases, to establish the reference values for exponents with higher precision we use an extended floating-point precision similar to [but not equivalent to] ``80-bit double'' or ``long double'' format.

\section{Numerical Error Propagation}
\label{section:results}
We identify five categories of numerical errors that we want to investigate:
\begin{itemize}
    \item compactification errors - the errors associated with representing the parametrizing functions on the compact interval $\tilde \rho \in [0, \tilde \rho_{\text{Max}}]$ rather than the entire non-negative real semiaxis;
    \item discretization errors - the errors connected to representing the parametrizing functions on a discrete grid;
    \item integration errors - the errors induced by performing the loop integrals numerically;
    \item stability-matrix-approximation errors - the errors coming from the finite-difference approximation of the derivatives defining the stability matrix;
    \item rounding errors - the errors of the floating-point arithmetic.
\end{itemize}
In this chapter, we present several analyses that quantify the effect of each of these errors on the precision of several leading \gls{rg} eigenvalues~$e_i$ and the anomalous dimension~$\eta$. In these analyses, we isolate specific categories of errors by examining how the precision of the critical exponents depends on various numerical parameters. For this purpose, we will often assume that some errors are uncorrelated, e.g. that the integration error is not impacted by a change in the grid spacing~$h_\rho$. With this assumption in mind, for each analysis, we identify a reference value~$e^{\text{ref}}$ for a critical exponent~$e$, with respect to which the numerical error is measured. The reference value is calculated with a significantly reduced error of the categories investigated in a given analysis. We emphasize, however, that this reference value is burdened by the other, uncorrelated kinds of numerical error.

All the results presented in this chapter were obtained for the three-dimensional $O(2)$ model. We performed analogous calculations for the $O(1)$ and $O(3)$ models and reached the same conclusions. For transparency, the results for these two models are presented in the repository in Ref.~\cite{Chlebicki2024Repository}. The details of numerical methods used for each of the analyses are summarized in \ref{appendix:numerical_details}.

\subsection{Compactification error}
Firstly, we investigate the error associated with restricting the parametrizing functions' domain to a compact set. We achieve that by tracking the critical exponents' errors with varying $\tilde \rho_{\text{Max}}$ while keeping $h_\rho = \tilde \rho_0 / 20$ fixed [$\tilde \rho_0$ denotes the minimum of the local potential]. This way, we separate the discretization error, controlled primarily by $h_\rho$, from the compactification error. In this calculation, specifying the reference values, with respect to which the errors are measured, is a challenging task because by increasing $\tilde \rho_{\text{Max}}$ we also increase the number of points, which might induce other kinds of errors, e.g. associated with the rising condition number of the stability matrix. The reference values were calculated with the extended floating-point precision with $\tilde \rho_{\text{Max}} = 3.5 \tilde{\rho}_0$. The coefficient $3.5$ was determined phenomenologically as the value for which the errors were minimized.

\begin{figure}
    \centering
    \begin{subfigure}{.49\textwidth}
        \includegraphics[width=0.9\textwidth]{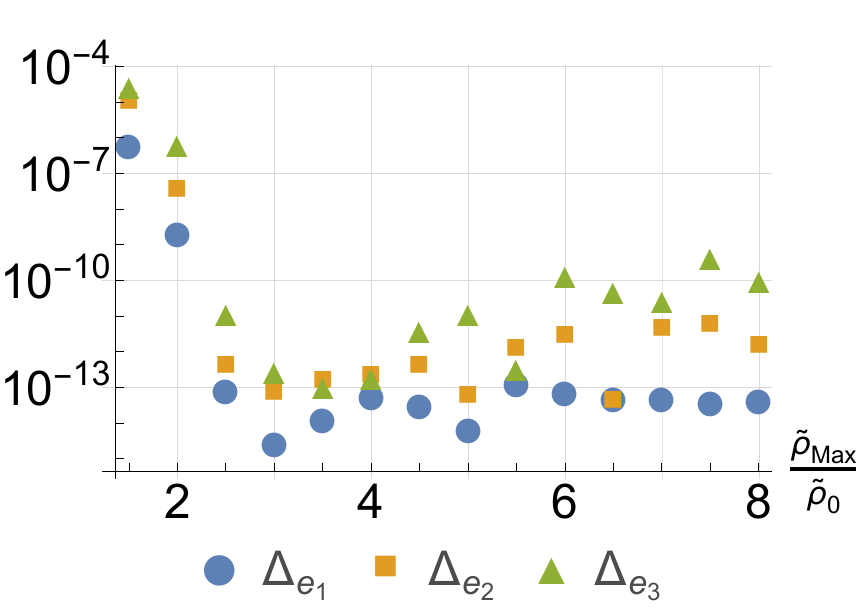}
        \subcaption{Results at the \gls{lpa} level.}
        \label{fig:compactification_error_lpa}
    \end{subfigure}
    \begin{subfigure}{.49\textwidth}
        \includegraphics[width=0.9\textwidth]{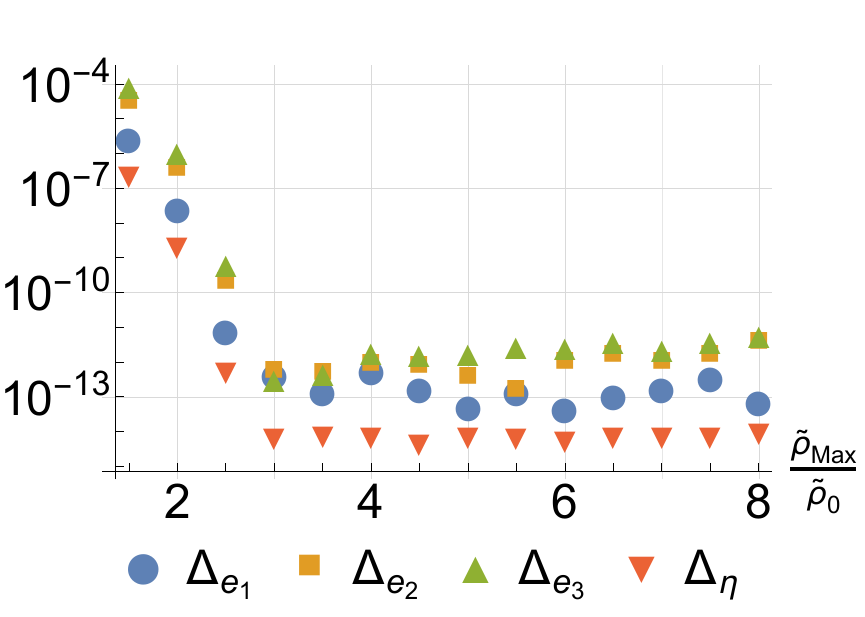}
        \subcaption{Results at the order $O(\partial^2)$.}
        \label{fig:compactification_error_de2}
    \end{subfigure}
    \caption{Precision of the critical exponents as a function of $\tilde \rho_{\text{Max}}$ for $h_\rho = \tilde \rho_0/20$. The reference values for each exponent were calculated for $\tilde \rho_{\text{Max}} = 3.5 \tilde \rho_0$ with the extended floating-point precision.}
    \label{fig:compactification_error}
\end{figure}

The results of this analysis at the \gls{lpa} and $O(\partial^2)$ orders of the \gls{de} are presented in Fig.~\ref{fig:compactification_error} as a function of $\tilde \rho_{\text{Max}}$. In both cases, the compactification error is relatively large at $\tilde \rho_{\text{Max}} = 1.5 \tilde \rho_0$ and drops precipitously with increasing $\tilde \rho_{\text{Max}}$ reaching a plateau at $\tilde \rho_{\text{Max}} \approx 3.5 \tilde \rho_0$. Beyond this value, a slight difference between the two cases arises. At the \gls{lpa} level, the plateau is very narrow and we can observe a slow deterioration of precision for large $\tilde \rho_{\text{Max}}$. At the order $O(\partial^2)$, no significant deterioration is observed and the plateau continues until $\tilde \rho_{\text{Max}} = 8 \tilde \rho_0$.

\subsection{Discretization error}
The primary parameter controlling the discretization error is the grid spacing $h_\rho$. Assuming the linear error propagation and disregarding the reduced accuracy of the derivative approximation at the borders, the error associated with discretization should be of order $O(h_\rho^{n-1})$ when the $n$-point approximation is employed. To verify the linear error propagation we track the error of the critical exponents as a function of the grid spacing using the $3$-, $5$-, and $7$-point derivative approximation. The reference values for this analysis are calculated using the $9$-point derivative approximation with the extended floating-point precision for $h_\rho = \tilde \rho_0/40$.

\begin{figure}
    \centering
    \begin{subfigure}{.49\textwidth}
        \includegraphics[width=0.9\textwidth]{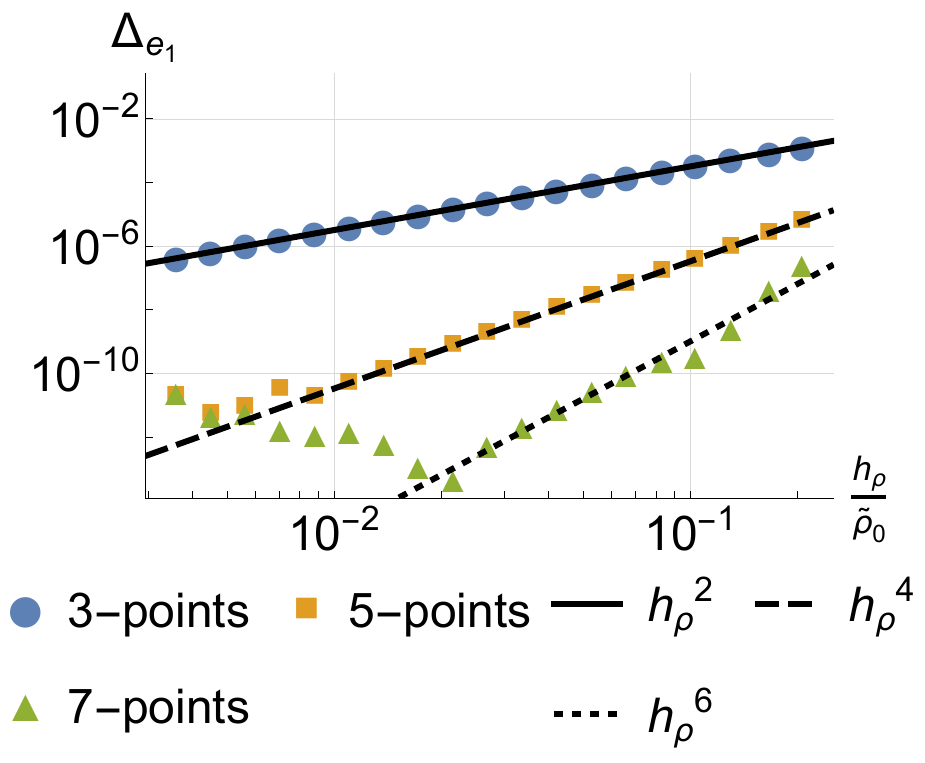}
        \subcaption{Leading eigenvalue $e_1$.}
    \end{subfigure}
    \begin{subfigure}{.49\textwidth}
        \includegraphics[width=0.9\textwidth]{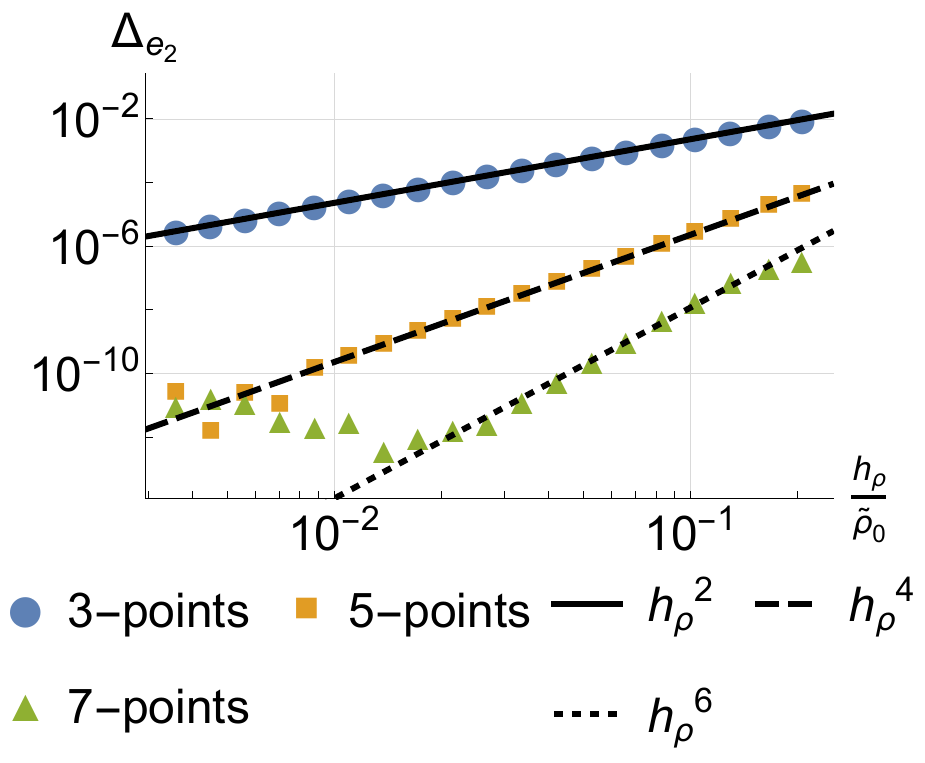}
        \subcaption{Subleading eigenvalue $e_2$.}
    \end{subfigure}
    \caption{Precision of the critical exponents as a function of $h_\rho $ for $\tilde \rho_{\text{Max}}= 3.5 \tilde \rho_0$ at the \gls{lpa} level. The reference values for each exponent were calculated with $9$-point derivative approximation for $\tilde h_\rho = \tilde \rho_0/40$ with the extended floating-point precision. Lines represent the best power-law fits to several trailing points with integer exponents.}
    \label{fig:discretization_lpa}
\end{figure}

\begin{figure}
    \centering
    \begin{subfigure}{.49\textwidth}
        \includegraphics[width=0.9\textwidth]{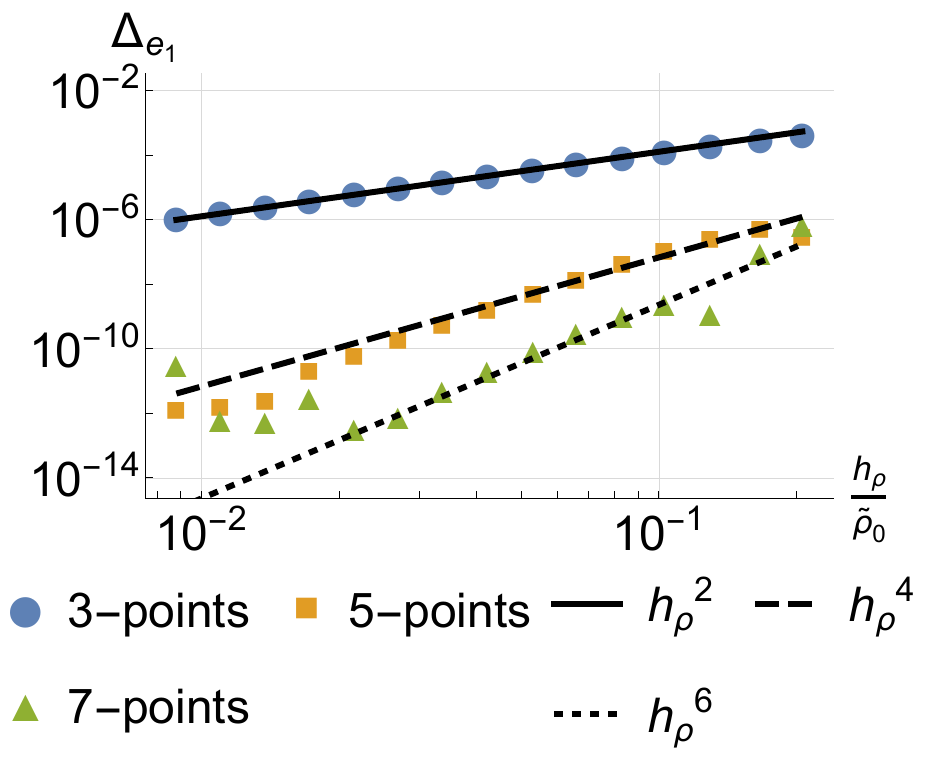}
        \subcaption{Leading eigenvalue $e_1$.}
    \end{subfigure}
    \begin{subfigure}{.49\textwidth}
        \includegraphics[width=0.9\textwidth]{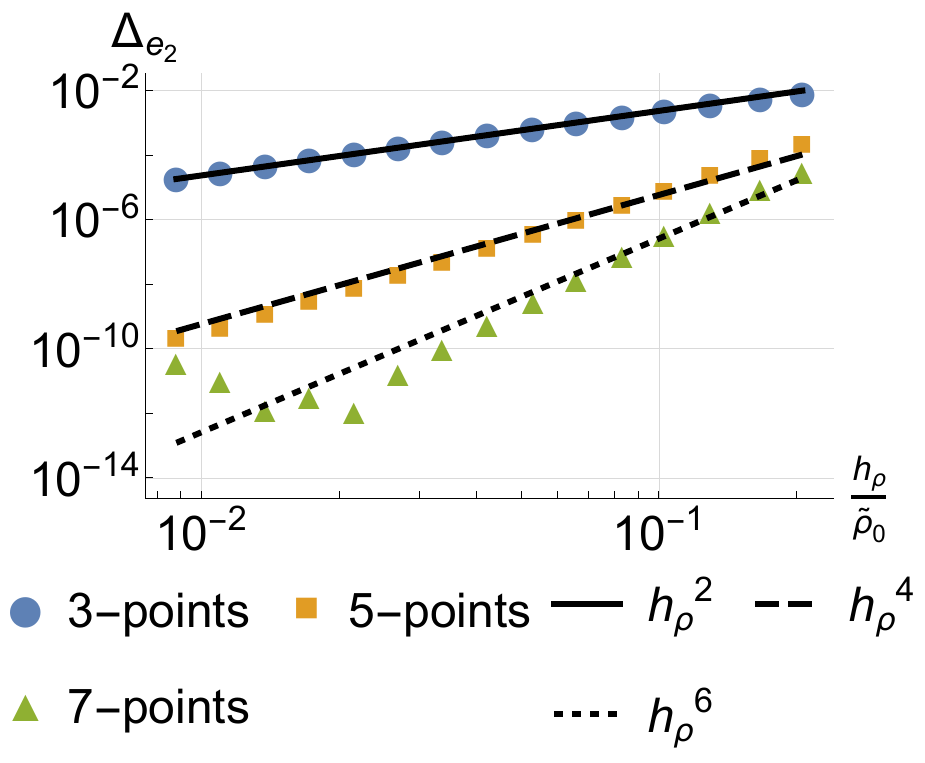}
        \subcaption{Subleading eigenvalue $e_2$.}
    \end{subfigure}
    \begin{subfigure}{.49\textwidth}
        \includegraphics[width=0.9\textwidth]{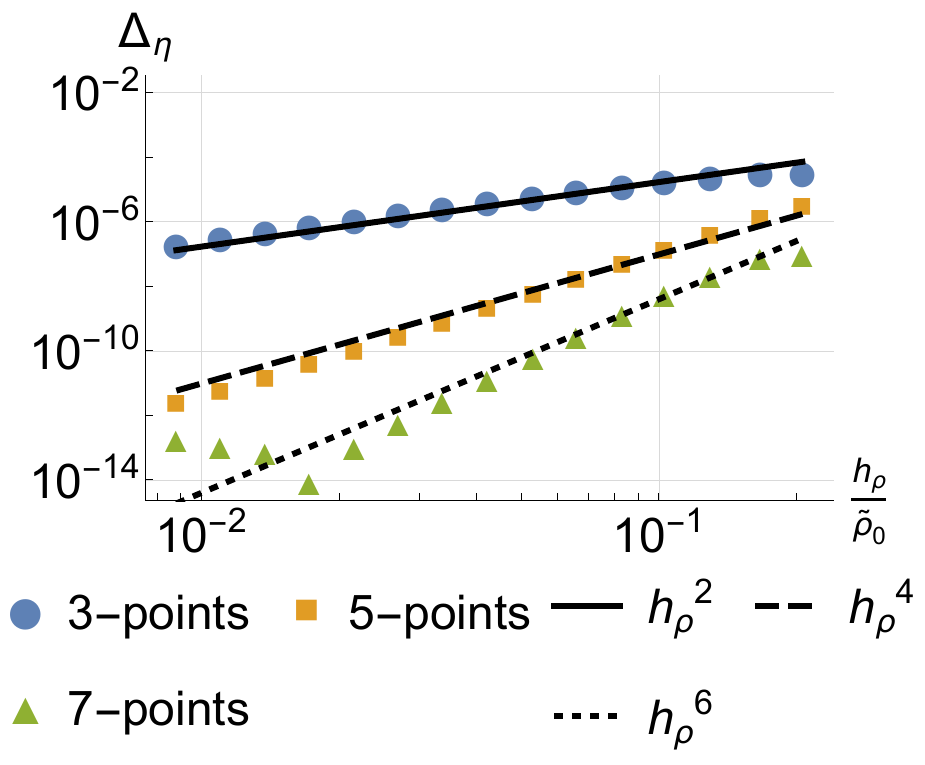}
        \subcaption{Anomalous dimension $\eta$.}
    \end{subfigure}
    \caption{Precision of the critical exponents as a function of $h_\rho $ for $\tilde \rho_{\text{Max}}= 3.5 \tilde \rho_0$ at the $O(\partial^2)$ order of the \gls{de}. The reference values for each exponent were calculated with $9$-point derivative approximation for $\tilde h_\rho = \tilde \rho_0/40$ with the extended floating-point precision. Lines represent the best power-law fits to several trailing points with integer exponents.}
    \label{fig:discretization_de2}
\end{figure}

In Fig.~\ref{fig:discretization_lpa}, we track the exponents' dependence on the grid spacing $h_\rho$ at the \gls{lpa} level for $\tilde \rho_{\text{Max}} = 3.5 \tilde \rho_0$ - the optimal value established in the previous analysis. All the data series, in the figure, tend to $0$ with $h_\rho \to 0$. This means that the estimates for the critical exponents converge to the same reference values irrespective of the discretization scheme. Moreover, the convergence rates are the same as the exponents controlling the corrections of the [bulk] finite-difference approximation for $\tilde \rho$~derivatives. For the $7$-point approximation, we can also observe the rounding errors overtake the discretization errors around $h_\rho \approx 0.02 \tilde \rho_0$. For lower values of $h_\rho$, the rounding errors lead to a slow decrease in accuracy. Essentially identical conclusions can be drawn from Fig.~\ref{fig:discretization_de2} presenting the results of analogous analysis at the order $O(\partial^2)$. This shows that, when properly set up, the discretization scheme employed in our calculations is convergent and can offer great accuracy.

\begin{figure}
    \centering
    \begin{subfigure}{.49\textwidth}
        \includegraphics[width=0.9\textwidth]{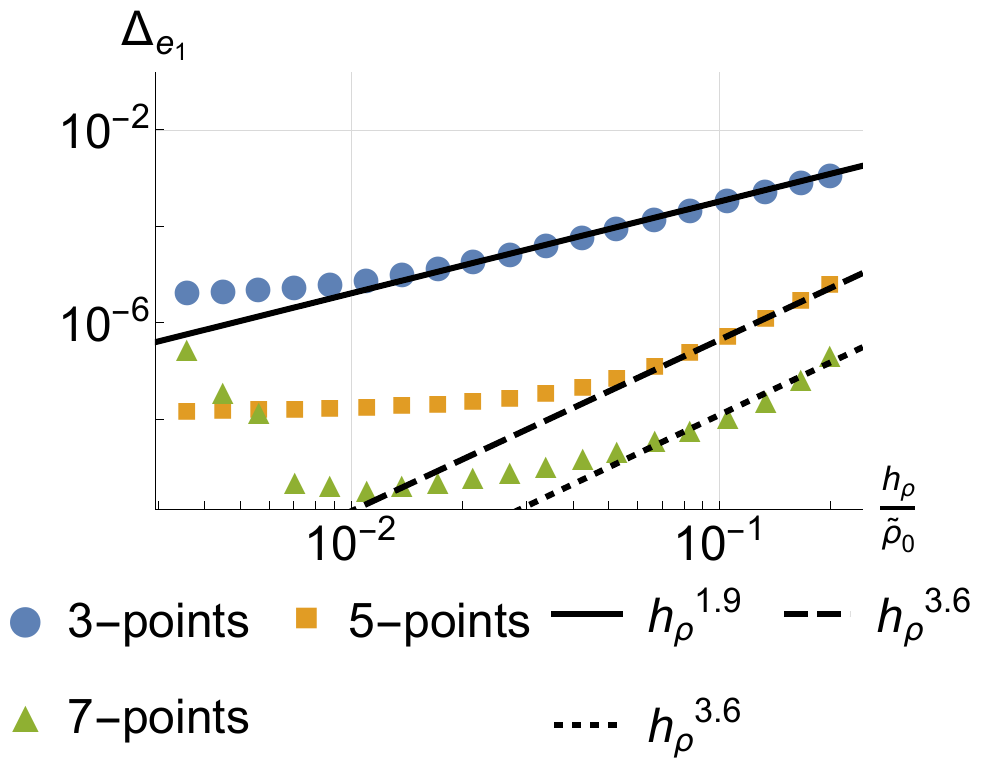}
        \subcaption{Leading eigenvalue $e_1$.}
    \end{subfigure}
    \begin{subfigure}{.49\textwidth}
        \includegraphics[width=0.9\textwidth]{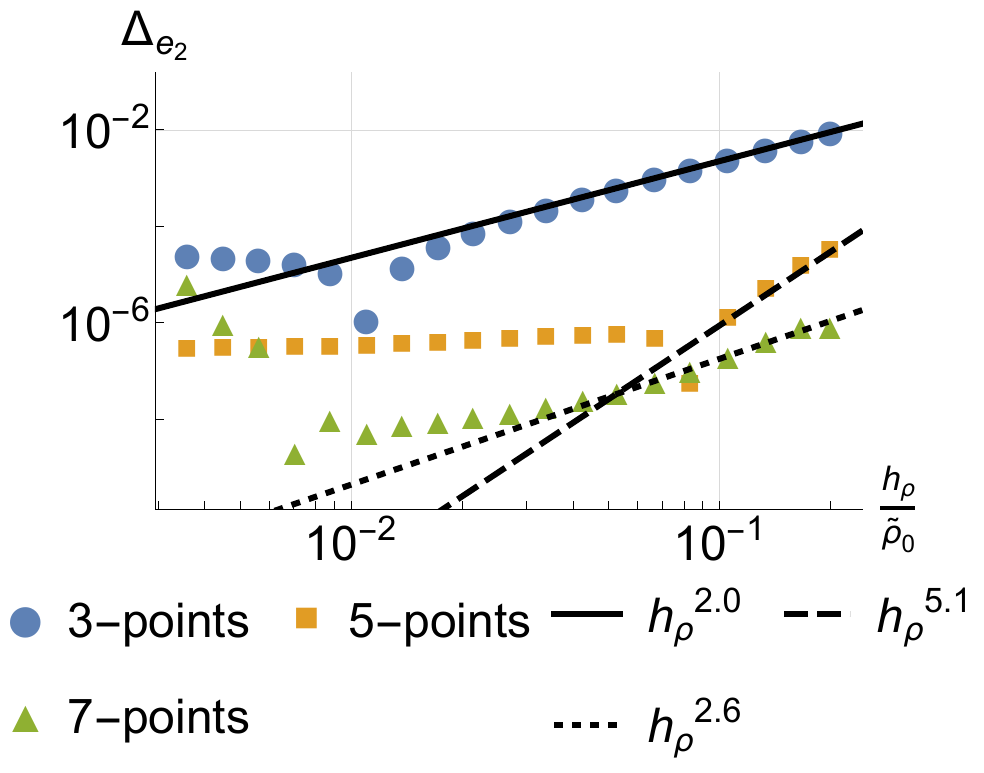}
        \subcaption{Subleading eigenvalue $e_2$.}
    \end{subfigure}
    \caption{Precision of the critical exponents as a function of $h_\rho $ for $\tilde \rho_{\text{Max}}= 2 \tilde \rho_0$ at the \gls{lpa} level. The reference values for each exponent were calculated with $9$-point derivative approximation for $\tilde h_\rho = \tilde \rho_0/40$ with the extended floating-point precision. Lines represent the best power-law fits to several trailing points.}
    \label{fig:discretization_lpa_lowrho}
\end{figure}

\begin{figure}
    \centering
    \begin{subfigure}{.49\textwidth}
        \includegraphics[width=0.9\textwidth]{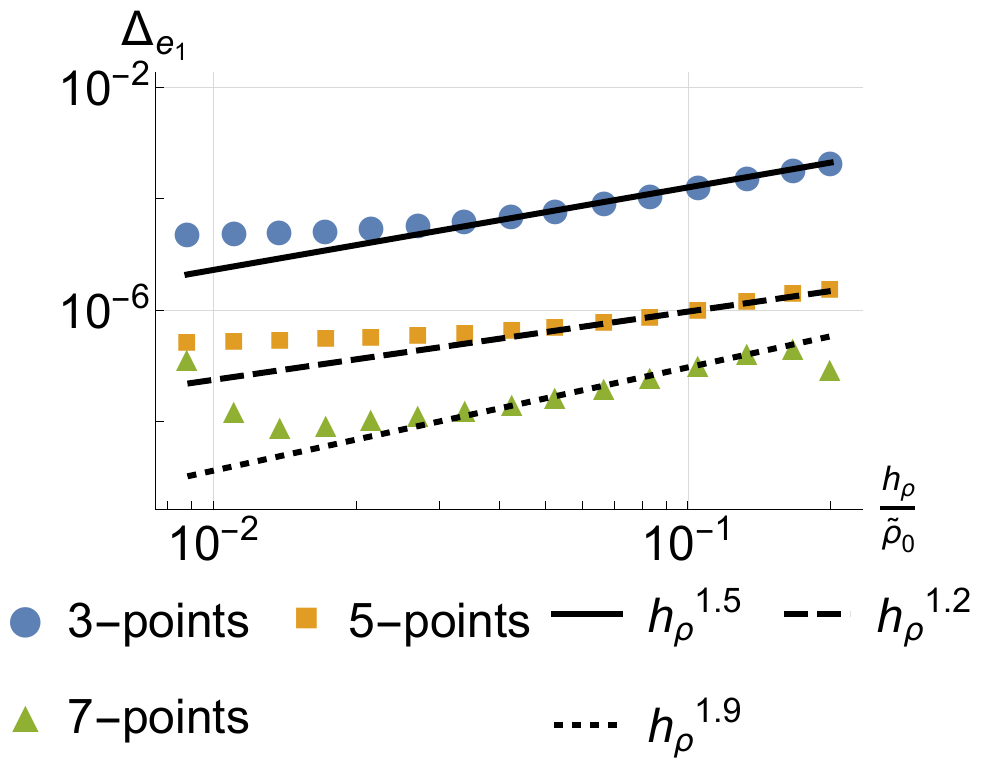}
        \subcaption{Leading eigenvalue $e_1$.}
    \end{subfigure}
    \begin{subfigure}{.49\textwidth}
        \includegraphics[width=0.9\textwidth]{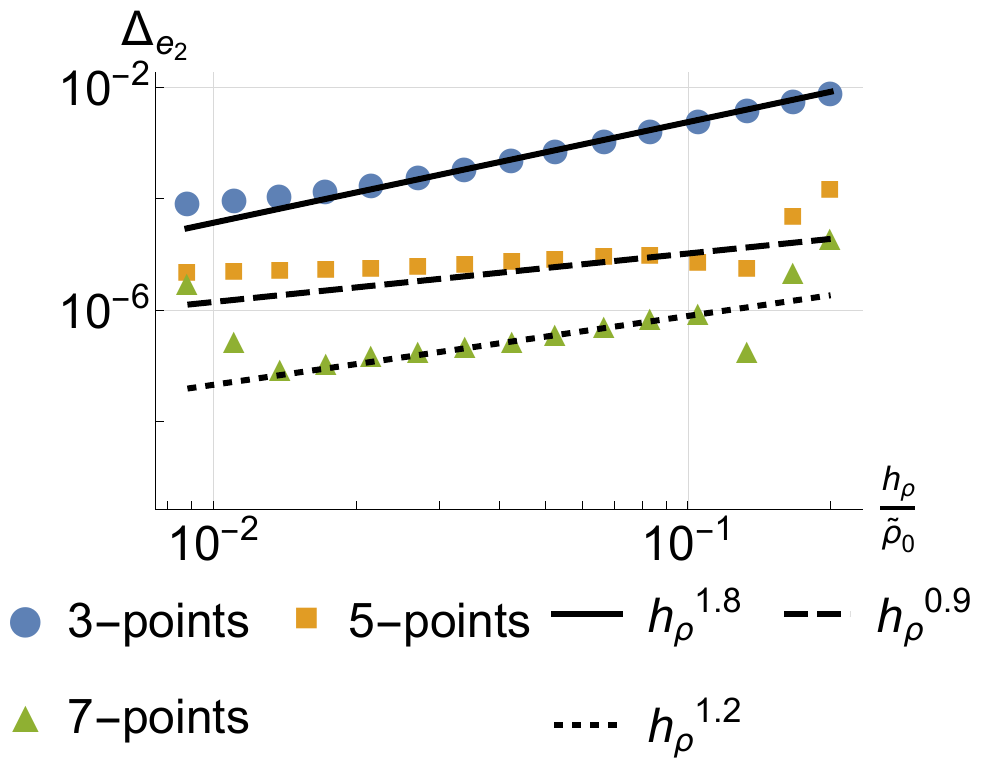}
        \subcaption{Subleading eigenvalue $e_2$.}
    \end{subfigure}
    \begin{subfigure}{.49\textwidth}
        \includegraphics[width=0.9\textwidth]{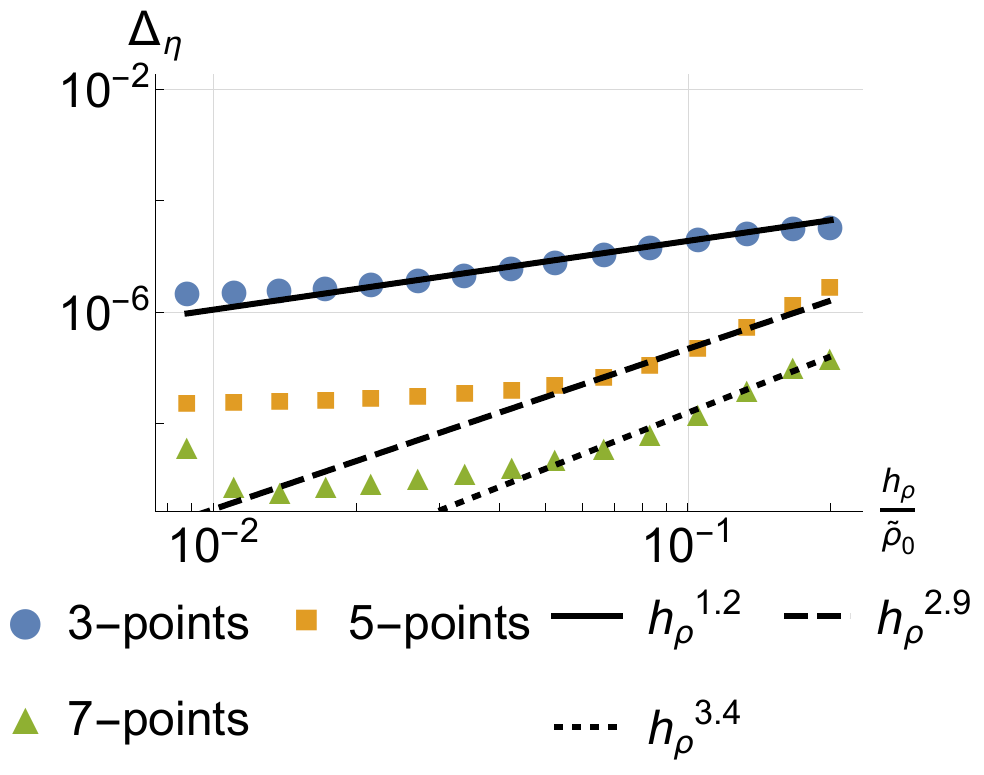}
        \subcaption{Anomalous dimension $\eta$.}
    \end{subfigure}
    \caption{Precision of the critical exponents as a function of $h_\rho $ for $\tilde \rho_{\text{Max}}= 2 \tilde \rho_0$ at the order $O(\partial^2)$. The reference values for each exponent were calculated with $9$-point derivative approximation for $\tilde h_\rho = \tilde \rho_0/40$ with the extended floating-point precision. Lines represent the best power-law fits to several trailing points.}
    \label{fig:discretization_de2_lowrho}
\end{figure}

We also performed an analysis similar to the one described above but with a significantly lower value of $\tilde \rho_{\text{Max}} = 2 \tilde \rho_0$. Its results are presented in Figs.~\ref{fig:discretization_lpa_lowrho} and \ref{fig:discretization_de2_lowrho} for the \gls{lpa} and $O(\partial^2)$ orders respectively. The overall precision and the convergence rates are significantly worse when compared to the results obtained for the optimal value of $\tilde \rho_{\text{Max}}$. More worryingly, instead of converging towards $0$, all data series in both figures display a plateau for a wide range of values of $h_\rho$. This implies that neither of the discretization schemes converges to the reference value. We verified that for each exponent each data series converges to a different value. This analysis showcases that the error induced by selecting a too small a value of $\tilde \rho_{\text{Max}}$ observed in Fig.~\ref{fig:compactification_error} is primarily caused by an impaired convergence of the discretization scheme.

\begin{figure}
    \centering
    \begin{subfigure}{.49\textwidth}
        \includegraphics[width=0.9\textwidth]{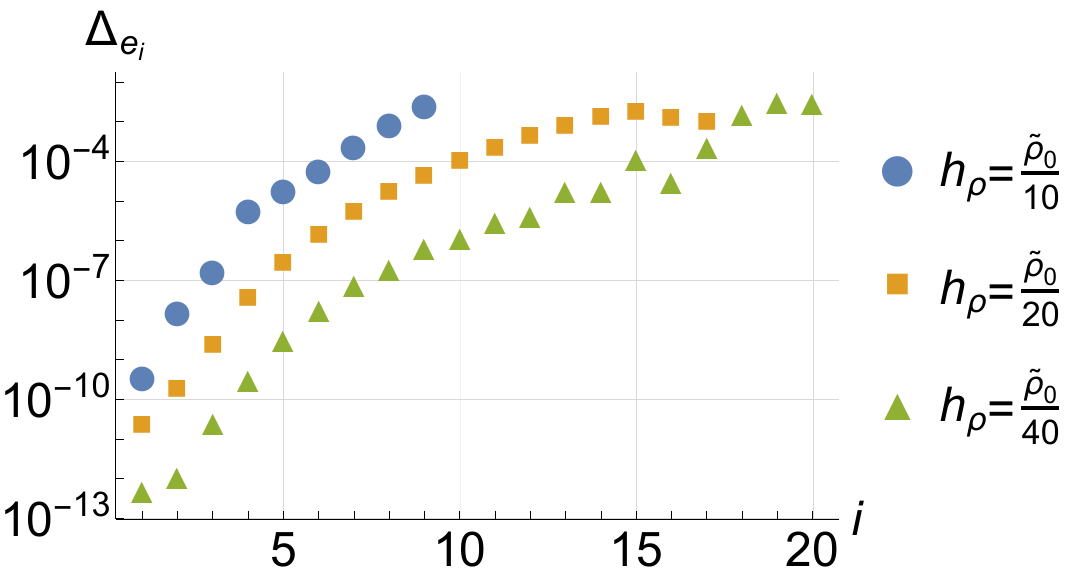}
        \subcaption{Results at the \gls{lpa} level.}
        \label{fig:eigenvalue_error_lpa}
    \end{subfigure}
    \begin{subfigure}{.49\textwidth}
        \includegraphics[width=0.9\textwidth]{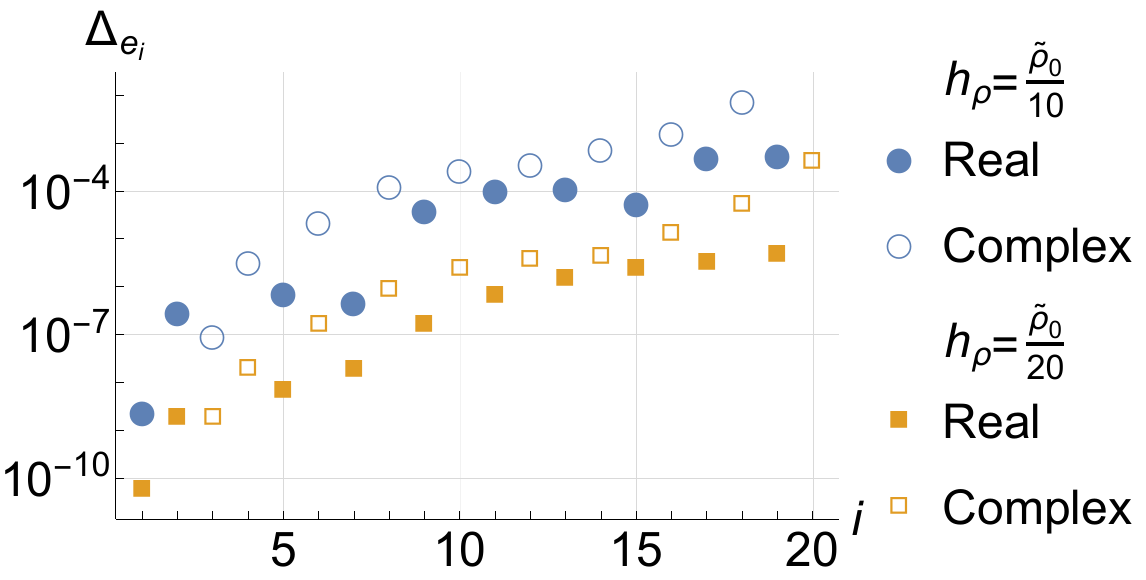}
        \subcaption{Results at the order $O(\partial^2)$.}
        \label{fig:eigenvalue_error_de2}
    \end{subfigure}
    \caption{Dependence of the eigenvalue error $\Delta_{e_i}$ on the eigenvalue index $i$ [sorted in descending order with respect to the real part] calculated for $\tilde \rho_{\text{Max}} = 3.5 \tilde \rho_0$ with different values of grid spacing $h_\rho$. The reference values for each exponent were calculated with the extended floating-point precision for $h_\rho = \tilde \rho_{0}/60$ at the \gls{lpa} level and $h_\rho = \tilde \rho_{0}/30$ at the order $O(\partial^2)$ using $7$-point derivative approximation.}
    \label{fig:eigenvalue_error}
\end{figure}

Finally, we compare the discretization accuracy between subsequent \gls{rg} eigenvalues. Fig.~\ref{fig:eigenvalue_error} shows the dependence of the eigenvalue error $\Delta_{e_i}$ on the eigenvalue index $i$ [sorted in descending order with respect to the real part]. At both the \gls{lpa} and the $O(\partial^2)$ orders of the \gls{de}, we can see an exponential decay of precision with each subsequent eigenvalue. This observation is particularly pertinent for studies of multicritical points where we are often interested in more than a few leading eigenvalues. 

In the figure, we marked a distinction between real and complex eigenvalues. We note that the stability matrix is real and non-symmetric so it can feature pairs of mutually conjugate complex eigenvalues. In the figure, such pairs have been reduced to a single point. Notably, complex eigenvalues are observed only at the $O(\partial^2)$ order of the \gls{de} and their precision is typically lower than that of similar real eigenvalues by approximately one order of magnitude. The origin and physical meaning of the complex eigenvalues are not well understood and require future clarification.

\subsection{Integration error}
\label{subsection:integration_error}
Let us now turn our attention to the numerical integration error. To be precise, we are not interested in the question of how to achieve a specific level of precision of numerical integrals. The answer to this question will be specific both to the investigated model and to the employed method. We ask the more general question about how the integration error propagates to errors in estimates of the critical exponents. To answer this question we calculate the critical exponents with progressively improving integral accuracy. On the technical level, we achieve this by performing the loop integrals with the Gauss-Legendre quadrature on the interval $\abs{\bm{q}} \in [0, 5.5]$ with the number of evaluation points $n$ varying between $5$ and $40$. For each value $n$, we define the integral error as:
\begin{equation}
    \Delta_I(n) = \max_{a, \tilde\rho_i}\abs{\tilde \beta^n_a(\tilde \rho_i) - \tilde \beta^{50}_a(\tilde \rho_i)}. \label{eq:integral_error}
\end{equation}
In the definition above, the maximum is taken over all parametrizing functions $a$ and all grid points $\tilde \rho_i$, $\tilde \beta^n_a$ denotes the dimensionless $\beta$ function for the function $a$ calculated with $n$-point integral approximation, and $\tilde \beta^{50}_a$ serves as a reference value. The $\beta$ functions in the definition \eqref{eq:integral_error} were evaluated at the \gls{frg} \gls{fp} solution obtained with $\tilde \rho_{\text{Max}} = 3.5 \tilde \rho_0$, $h_\rho = \tilde \rho_0/20$ and $50$-point integral approximation. Since the Litim regulator allows for analytical integration at the \gls{lpa} level, it would be pointless to investigate the integration error in this setting. Therefore, in this analysis, we employ the exponential regulator both at the \gls{lpa} and $O(\partial^2)$ orders.

\begin{figure}
    \centering
    \begin{subfigure}{.49\textwidth}
        \includegraphics[width=0.9\textwidth]{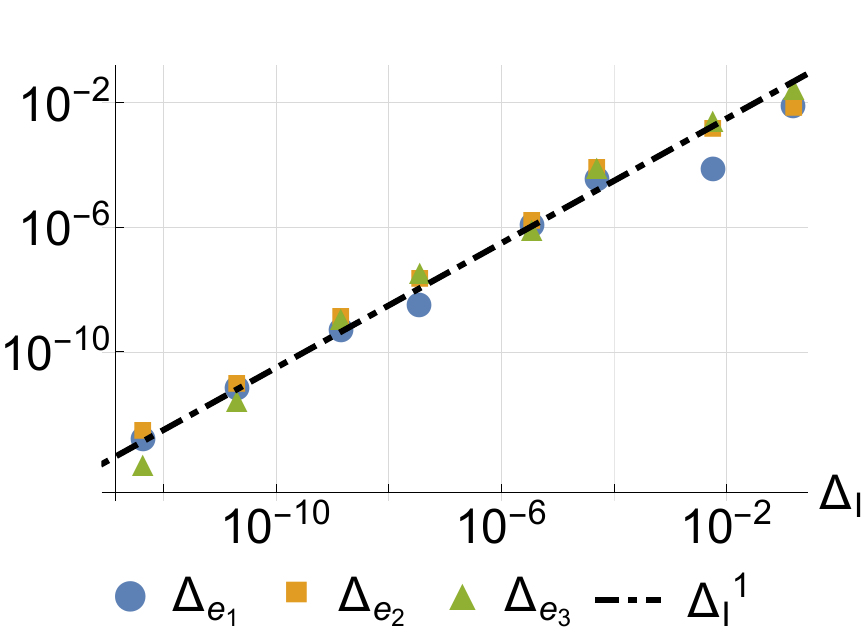}
        \subcaption{Results at the \gls{lpa} level.}
        \label{fig:integral_error_lpa}
    \end{subfigure}
    \begin{subfigure}{.49\textwidth}
        \includegraphics[width=0.9\textwidth]{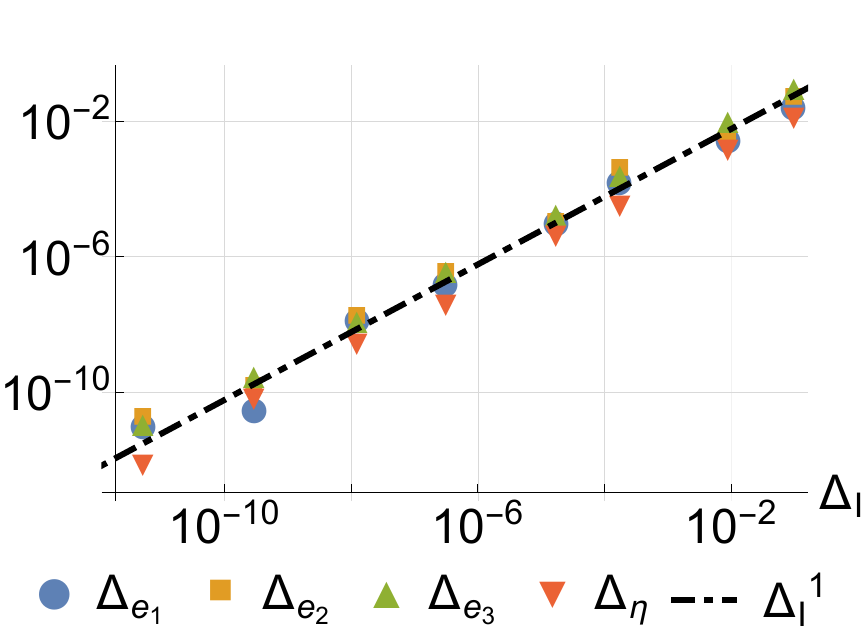}
        \subcaption{Results at the order $O(\partial^2)$.}
        \label{fig:integral_error_de2}
    \end{subfigure}
    \caption{Precision of the critical exponents as a function of integral precision $\Delta_I$. The lines denote the best linear fit to several trailing points. In the calculation, we employed the $7$-point approximation for the $\tilde \rho$~derivatives with $\tilde \rho_{\text{Max}} = 3.5 \tilde \rho_0$ and $h_\rho = \tilde \rho_0 / 20$.}
    \label{fig:integral_error}
\end{figure}

In Fig.~\ref{fig:integral_error}, we chart the precision of the critical exponents against the precision of the loop integrals. The figure clearly shows that the two errors are directly proportional. Moreover, the errors of the critical exponents are of the same order of magnitude as the integration errors $\Delta_I$. This means that, when implementing the numerical procedures for the loop integral, we do not need to test their accuracy on the critical exponents which would require a significant numerical effort; it is sufficient to test their accuracy directly on the $\beta$ functions.

\subsection{Stability matrix approximation error}
\label{subsection:stability_matrix_approximation_error}
Finally, we examine the finite-difference approximation of the stability matrix. We calculate the eigenvalues of the stability matrix constructed with $3$-, $5$-, and $7$-point approximations [see Eq.~\eqref{eq:finite_difference_approximation}] for a wide range of values for the perturbation parameter $\epsilon$ and compare them to the eigenvalues of analytically calculated stability matrix. 

\begin{figure}
    \centering
    \begin{subfigure}{.49\textwidth}
        \includegraphics[width=0.9\textwidth]{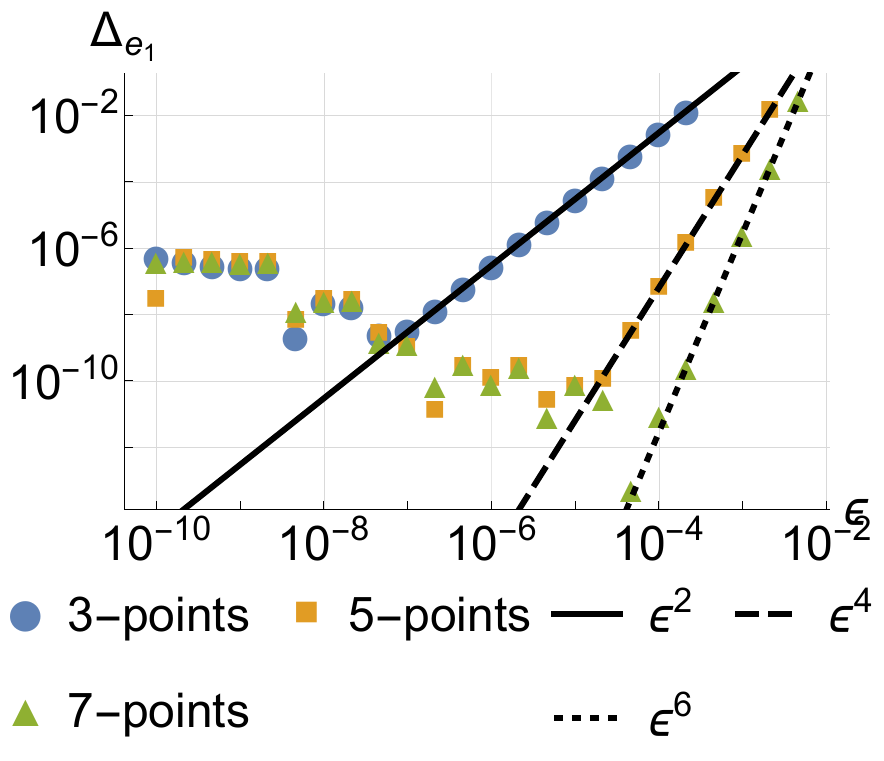}
        \subcaption{Leading eigenvalue $e_1$.}
    \end{subfigure}
    \begin{subfigure}{.49\textwidth}
        \includegraphics[width=0.9\textwidth]{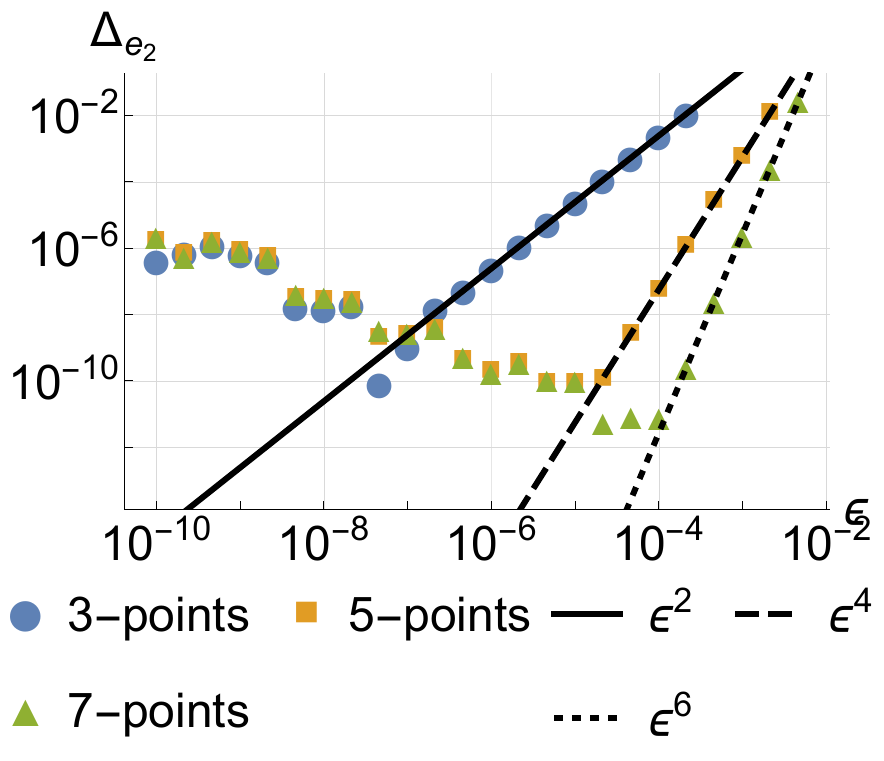}
        \subcaption{Subleading eigenvalue $e_2$.}
    \end{subfigure}
    \caption{Precision of the two leading \gls{rg} eigenvalues as a function of the perturbation $\epsilon$ used in the stability-matrix finite-difference approximation calculated at the \gls{lpa} level. Lines represent the best power-law fits to several trailing points with integer exponents. In the calculation, we employed the $7$-point approximation for the $\tilde \rho$~derivatives with $\tilde \rho_{\text{Max}} = 3.5 \tilde \rho_0$ and $h_\rho = \tilde \rho_0 / 20$. The reference values were extracted from the analytically constructed stability matrix.}
    \label{fig:stability_matrix_lpa}
\end{figure}

\begin{figure}
    \centering
    \begin{subfigure}{.49\textwidth}
        \includegraphics[width=0.9\textwidth]{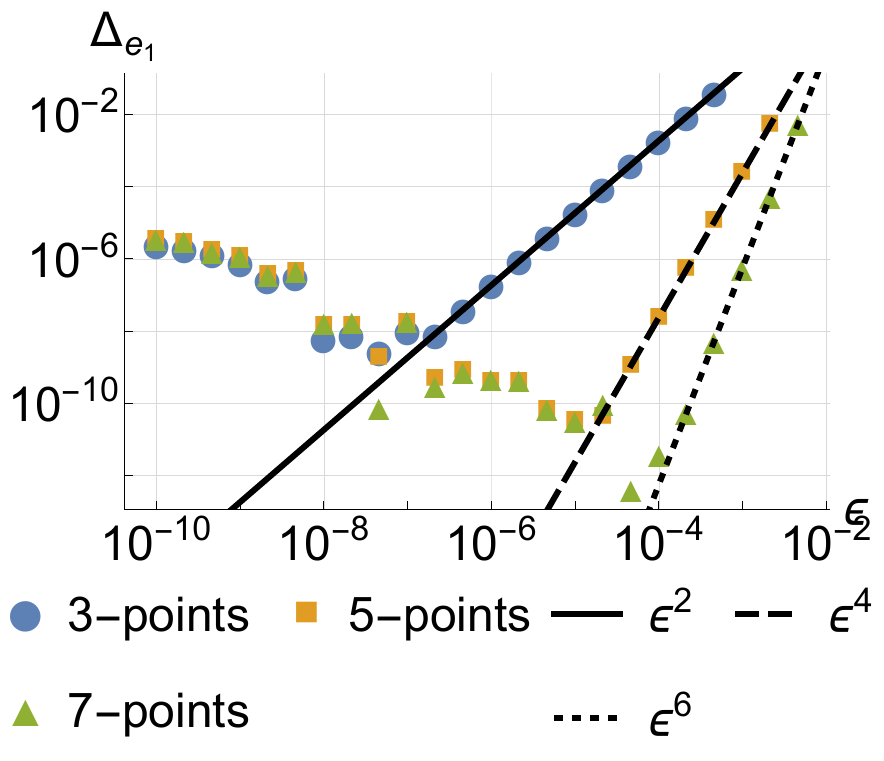}
        \subcaption{Leading eigenvalue $e_1$.}
    \end{subfigure}
    \begin{subfigure}{.49\textwidth}
        \includegraphics[width=0.9\textwidth]{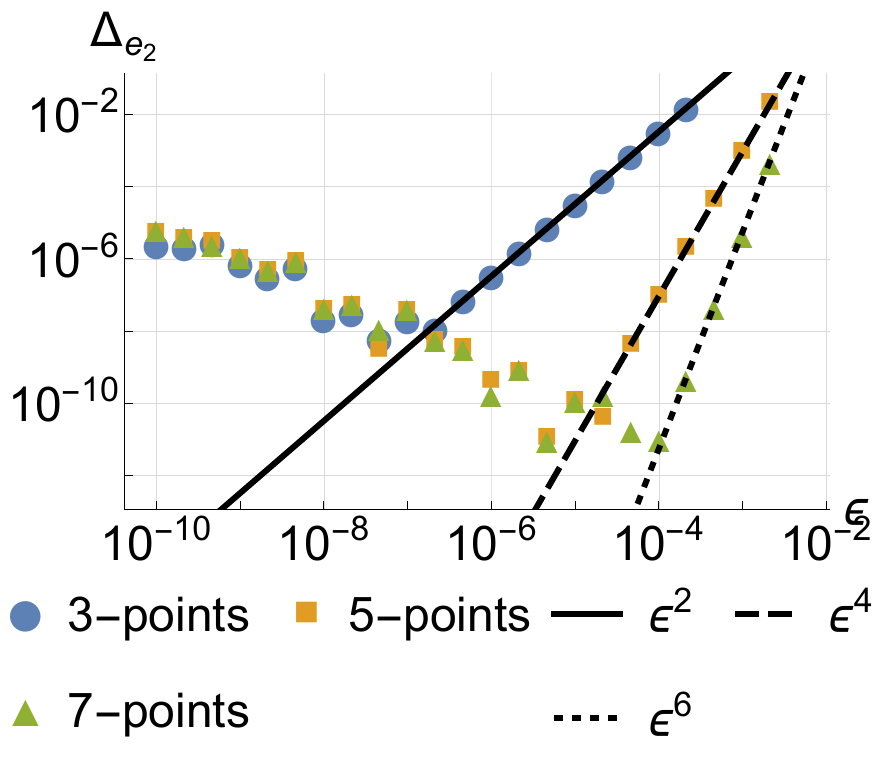}
        \subcaption{Subleading eigenvalue $e_2$.}
    \end{subfigure}
    \caption{Precision of the two leading \gls{rg} eigenvalues as a function of the perturbation $\epsilon$ used in the stability-matrix finite-difference approximation calculated at the order $O(\partial^2)$. Calculation were performed with $\tilde \rho_{\text{Max}} = 3.5 \tilde \rho_0$ and $h_\rho = \tilde \rho_0 / 20$. The reference values were extracted from the analytically constructed stability matrix. Lines represent the best power-law fits to several trailing points with integer exponents.}
    \label{fig:stability_matrix_de2}
\end{figure}

Fig.~\ref{fig:stability_matrix_lpa} presents the results of this analysis at the \gls{lpa} level. For large values of~$\epsilon$, we observe that the approximation is convergent with the expected rate of $n-1$ for $n$-point approximation. For small enough $\epsilon$, each approximation breaks down as the rounding errors overtake the finite-difference-approximation errors. Interestingly, for all three approximations, the rounding errors all follow the exact same line $\sim \epsilon^{-1}$ with just a few random outliers \footnote{The outliers can be expected, as sometimes despite the rounding error the approximate value can randomly land close to the exact value. Importantly, there are no outliers above the $\epsilon^{-1}$ line.}. Fig.~\ref{fig:stability_matrix_de2} shows the results of analogous calculation at the order $O(\partial^2)$. Remarkably, the results for two orders of the \gls{de} are almost identical.

\section{Conclusion}
\label{section:conclusion}
In this work, we investigated the numerical accuracy of the stability-matrix approach to the functional renormalization group. We thoroughly examined the common numerical implementation of this scheme and identified major sources of numerical errors. Subsequently, we performed several tests to quantify the magnitude of these errors and verify the convergence of our numerical scheme. 

Our calculations clearly show that the stability-matrix approach with the grid representation is a robust scheme allowing for remarkably precise calculations. Performing the calculations in the 64-bit floating-point representation we can calculate the five leading \gls{rg} eigenvalues with the numerical error smaller than $10^{-8}$, the value far below the typical error of the \gls{de} at this order lying between $10^{-2}$ and $10^{-4}$ \cite{DePolsi2020}. We emphasize, that such accuracy can only be achieved when all the parameters of the numerical implementation are properly tuned.

Interestingly, the numerical errors obtained at the order $O(\partial^2)$ are essentially identical to those calculated at the \gls{lpa} level, despite a three-times difference in the number of parameters and a wide difference in complexity of the $\beta$ functions. We can, thus, reasonably expect that the order of magnitude of the numerical errors characterizing higher orders of the \gls{de} should remain roughly the same. We stress that, if needed, the numerical errors can be reduced even further by adopting a more precise floating-point representation.

Most of the calculations presented in this work, show that the employed numerical schemes converge with the theoretically predicted convergence rate. There is, however, one important exception to this rule. Selecting too small a value for the upper bound of the $\tilde \rho$ grid $\tilde \rho_{\text{Max}}$ strongly impairs the convergence with respect to the grid spacing~$h_\rho$. We have shown that imposing $\tilde \rho_{\text{Max}} < 3.5 \tilde \rho_0$ leads to a significant deterioration of the numerical accuracy. We note that increasing $\tilde \rho_{\text{Max}}$ beyond $3.5 \tilde \rho_0$ [with fixed $h_\rho$] has essentially no effect on the accuracy but quickly increases the numerical burden.

We emphasize that all calculations for this work were performed in the relatively simple three-dimensional $O(N)$ models. While we can expect the general conclusions to hold more broadly, specific error estimates should be recalculated for each case. The rounding errors are particularly strongly implementation-dependent and have to be carefully assessed. We suggest the methods presented in this work as a convenient tool for benchmarking the accuracy of numerical implementations of the \gls{frg} and tuning the values of numerical parameters.

%We conclude with a general remark. By writing this work we aim to invite more attention and scrutiny to numerical implementations of our methodology. This topic is hugely important; to a numericist, a numerical implementation is no different than an experimental setup to an experimentalist. Yet, this part of the research process often remains overlooked. Many studies do not offer detailed descriptions of the methods employed and almost none of them describe the testing performed to ensure the reliability of the results. This makes it needlessly difficult to reproduce the previous results and is contrary to the principles of open science. In our opinion, the only real way for the numerical results to be scrutinized is to share the programs used in the calculation in an open-source format.

\section*{Acknowledgements}
We are grateful to Paweł Jakubczyk for the discussions as well as for reading the early version of the manuscript and very useful comments.  We thank Carlos A. Sanch\'ez Villalobos for his useful comments on the early version of the manuscript.

We acknowledge funding from the Polish National Science Center via grant 2017/26/E/ST3/00211 and from the University of Warsaw via IDUB Programme, Action IV.4.1. \textit{A complex programme of support for UW PhD students}. 

\appendix
\begin{landscape}
\section{Details of numerical calculations}
\label{appendix:numerical_details}
\begin{table}[h]
    \centering \resizebox{1.2\textwidth}{!}{%
    \begin{tabular}{ccccccccc}\bottomrule
     Calculation    & DE order & $\tilde \rho_{\text{Max}}$ & $h_\rho$ & DA & IR & Integral & SM & FPP  \\ \midrule
     Compactification error [Fig.~\ref{fig:compactification_error_lpa}] & \multirow{2}{*}{LPA} & Varied & \multirow{2}{*}{$\tilde \rho_0 / 20$} & \multirow{2}{*}{$7$-point} & \multirow{2}{*}{$[0, 1]$} & \multirow{2}{*}{Analytical} & \multirow{2}{*}{Analytical} & 64-bit double \\ 
     Reference value &  & $3.5 \tilde \rho_0$ &  &  &  &  &  & 80-bit double \\  \midrule
     
     Compactification error [Fig.~\ref{fig:compactification_error_de2}] & \multirow{2}{*}{$O(\partial^2)$} & Varied & \multirow{2}{*}{$\tilde \rho_0 / 20$} & \multirow{2}{*}{$7$-point} & \multirow{2}{*}{$[0, 5]$} & \multirow{2}{*}{35-point GL} & \multirow{2}{*}{Analytical} & 64-bit double \\ 
     Reference value &  & $3.5 \tilde \rho_0$ & & & & & & 80-bit double \\  \midrule
     
     Discretization error [Fig.~\ref{fig:discretization_lpa}] & \multirow{2}{*}{LPA} & \multirow{2}{*}{$3.5 \tilde \rho_0$} & Varied & Varied & \multirow{2}{*}{$[0, 1]$} & \multirow{2}{*}{Analytical} & \multirow{2}{*}{Analytical} & 64-bit double \\
     Reference value & & & $\tilde \rho_0 / 40$ & $9$-point & & & & 80-bit double \\ \midrule
     
     Discretization error [Fig.~\ref{fig:discretization_de2}] & \multirow{2}{*}{$O(\partial^2)$} & \multirow{2}{*}{$3.5 \tilde \rho_0$} & Varied & Varied & \multirow{2}{*}{$[0, 5]$} & \multirow{2}{*}{35-point GL} & \multirow{2}{*}{Analytical} & 64-bit double \\
     Reference value & & & $\tilde \rho_0 / 40$ & $9$-point & & & & 80-bit double \\ \midrule
     
     Discretization error [Fig.~\ref{fig:discretization_lpa_lowrho}] & \multirow{2}{*}{LPA} & \multirow{2}{*}{$2 \tilde \rho_0$} & Varied & Varied & \multirow{2}{*}{$[0, 1]$} & \multirow{2}{*}{Analytical} & \multirow{2}{*}{Analytical} & 64-bit double \\
     Reference value & & & $\tilde \rho_0 / 40$ & $9$-point & & & & 80-bit double \\ \midrule
     
     Discretization error [Fig.~\ref{fig:discretization_de2_lowrho}] & \multirow{2}{*}{$O(\partial^2)$} & \multirow{2}{*}{$2 \tilde \rho_0$} & Varied & Varied & \multirow{2}{*}{$[0, 5]$} & \multirow{2}{*}{35-point GL} & \multirow{2}{*}{Analytical} & 64-bit double \\
     Reference value & & & $\tilde \rho_0 / 40$ & $9$-point & & & & 80-bit double \\ \midrule

     Discretization error [Fig.~\ref{fig:eigenvalue_error_lpa}] & \multirow{2}{*}{LPA} & \multirow{2}{*}{$3.5 \tilde \rho_0$} & Varied & \multirow{2}{*}{$7$-point} & \multirow{2}{*}{$[0, 1]$} & \multirow{2}{*}{Analytical} & \multirow{2}{*}{Analytical} & 64-bit double  \\
     Reference values & & & $\tilde \rho_0 / 60$ & & & & & 80-bit double \\ \midrule

     Discretization error [Fig.~\ref{fig:eigenvalue_error_de2}] & \multirow{2}{*}{$O(\partial^2)$} & \multirow{2}{*}{$3.5 \tilde \rho_0$} & Varied & \multirow{2}{*}{$7$-point} & \multirow{2}{*}{$[0, 5]$} & \multirow{2}{*}{35-point GL} & \multirow{2}{*}{Analytical} & 64-bit double  \\
     Reference values & & & $\tilde \rho_0 / 30$ & & & & & 80-bit double \\ \midrule
     
     Integration error [Fig.~\ref{fig:integral_error}] & \multirow{2}{*}{LPA/$O(\partial^2)$} & \multirow{2}{*}{$3.5 \tilde \rho_0$} & \multirow{2}{*}{$\tilde \rho_0/20$} & \multirow{2}{*}{$7$-point} & \multirow{2}{*}{$[0, 5.5]$} & Varied GL & \multirow{2}{*}{Analytical} & 64-bit double \\
     Reference values & & & & & & 50-point GL & & 80-bit double \\ \midrule
     
     Stability-matrix error [Figs.~\ref{fig:stability_matrix_lpa}] & \multirow{2}{*}{LPA} &  \multirow{2}{*}{$3.5 \tilde \rho_0$} &  \multirow{2}{*}{$\tilde \rho_0/20$} &  \multirow{2}{*}{$7$-point} &  \multirow{2}{*}{$[0, 1]$} &  \multirow{2}{*}{Analytical} & Finite difference [Varied] &  \multirow{2}{*}{64-bit double} \\
     Reference values & & & & & & & Analytical & \\ \midrule
     
     Stability-matrix error [Figs.~\ref{fig:stability_matrix_de2}] & \multirow{2}{*}{$O(\partial^2)$} &  \multirow{2}{*}{$3.5 \tilde \rho_0$} &  \multirow{2}{*}{$\tilde \rho_0/20$} &  \multirow{2}{*}{$7$-point} &  \multirow{2}{*}{$[0, 5]$} &  \multirow{2}{*}{35-point GL} & Finite difference [Varied] &  \multirow{2}{*}{64-bit double} \\
     Reference values & & & & & & & Analytical & \\ \bottomrule
    \end{tabular}}
    \caption{Numerical parameters of all calculations performed in this study. $\tilde \rho_{\text{Max}}$ denotes the upper bound of the $\tilde \rho$ grid, and $h_\rho$ denotes the grid spacing. DA stands for the $\tilde \rho$ derivative approximation, IR - for the loop integration range [with respect to~$q = \abs{\bm q}$], GL - for the Gauss-Legendre quadrature, SM - for the method of calculating the derivatives defining the stability matrix, and FPP~-~for the floating-point precision.}
    \label{tab:calculation_details}
\end{table}
\end{landscape}
\section{Discrete derivative operators}
\label{appendix:discrete_derivatives}
In Sec.~\ref{section:numerical_methods}, we discussed the finite-difference approximations for derivatives in the $\tilde \rho$-discretization scheme and provided expressions for the $3$-point discrete derivative operators $D^1_3$ and $D^2_3$ approximating $\partial_{\tilde \rho}$ and $\partial^2_{\tilde\rho}$ operators respectively. Below, we present the $5$-point operators $D^1_5$ and $D^2_5$ for $N_\rho=6$, the $7$-point operators $D^1_7$ and $D^2_7$ for $N_\rho=8$, and the $9$-point operators $D^1_9$ and $D^2_9$ for $N_\rho=10$:
\begin{subequations}
\begin{align}
&D_5^1= \frac{1}{h_\rho}\left(
\begin{array}{ccccccc}
 -\frac{25}{12} & 4 & -3 & \frac{4}{3} & -\frac{1}{4} & 0 & 0 \\
 -\frac{1}{4} & -\frac{5}{6} & \frac{3}{2} & -\frac{1}{2} & \frac{1}{12} & 0 & 0 \\
 \frac{1}{12} & -\frac{2}{3} & 0 & \frac{2}{3} & -\frac{1}{12} & 0 & 0 \\
 0 & \frac{1}{12} & -\frac{2}{3} & 0 & \frac{2}{3} & -\frac{1}{12} & 0 \\
 0 & 0 & \frac{1}{12} & -\frac{2}{3} & 0 & \frac{2}{3} & -\frac{1}{12} \\
 0 & 0 & -\frac{1}{12} & \frac{1}{2} & -\frac{3}{2} & \frac{5}{6} & \frac{1}{4} \\
 0 & 0 & \frac{1}{4} & -\frac{4}{3} & 3 & -4 & \frac{25}{12} \\
\end{array}
\right), \\
&D_5^2 = \frac{1}{h_\rho^2}\left(
\begin{array}{ccccccc}
 \frac{35}{12} & -\frac{26}{3} & \frac{19}{2} & -\frac{14}{3} & \frac{11}{12} & 0 & 0 \\
 \frac{11}{12} & -\frac{5}{3} & \frac{1}{2} & \frac{1}{3} & -\frac{1}{12} & 0 & 0 \\
 -\frac{1}{12} & \frac{4}{3} & -\frac{5}{2} & \frac{4}{3} & -\frac{1}{12} & 0 & 0 \\
 0 & -\frac{1}{12} & \frac{4}{3} & -\frac{5}{2} & \frac{4}{3} & -\frac{1}{12} & 0 \\
 0 & 0 & -\frac{1}{12} & \frac{4}{3} & -\frac{5}{2} & \frac{4}{3} & -\frac{1}{12} \\
 0 & 0 & -\frac{1}{12} & \frac{1}{3} & \frac{1}{2} & -\frac{5}{3} & \frac{11}{12} \\
 0 & 0 & \frac{11}{12} & -\frac{14}{3} & \frac{19}{2} & -\frac{26}{3} & \frac{35}{12} \\
\end{array}
\right), 
\end{align}
\begin{align}
&D_7^1= \frac{1}{h_\rho}\left(
\begin{array}{ccccccccc}
 -\frac{49}{20} & 6 & -\frac{15}{2} & \frac{20}{3} & -\frac{15}{4} & \frac{6}{5} & -\frac{1}{6} & 0 & 0 \\
 -\frac{1}{6} & -\frac{77}{60} & \frac{5}{2} & -\frac{5}{3} & \frac{5}{6} & -\frac{1}{4} & \frac{1}{30} & 0 & 0 \\
 \frac{1}{30} & -\frac{2}{5} & -\frac{7}{12} & \frac{4}{3} & -\frac{1}{2} & \frac{2}{15} & -\frac{1}{60} & 0 & 0 \\
 -\frac{1}{60} & \frac{3}{20} & -\frac{3}{4} & 0 & \frac{3}{4} & -\frac{3}{20} & \frac{1}{60} & 0 & 0 \\
 0 & -\frac{1}{60} & \frac{3}{20} & -\frac{3}{4} & 0 & \frac{3}{4} & -\frac{3}{20} & \frac{1}{60} & 0 \\
 0 & 0 & -\frac{1}{60} & \frac{3}{20} & -\frac{3}{4} & 0 & \frac{3}{4} & -\frac{3}{20} & \frac{1}{60} \\
 0 & 0 & \frac{1}{60} & -\frac{2}{15} & \frac{1}{2} & -\frac{4}{3} & \frac{7}{12} & \frac{2}{5} & -\frac{1}{30} \\
 0 & 0 & -\frac{1}{30} & \frac{1}{4} & -\frac{5}{6} & \frac{5}{3} & -\frac{5}{2} & \frac{77}{60} & \frac{1}{6} \\
 0 & 0 & \frac{1}{6} & -\frac{6}{5} & \frac{15}{4} & -\frac{20}{3} & \frac{15}{2} & -6 & \frac{49}{20} \\
\end{array}
\right), \\
&D_7^2 = \frac{1}{h_\rho^2}\left(
\begin{array}{ccccccccc}
 \frac{203}{45} & -\frac{87}{5} & \frac{117}{4} & -\frac{254}{9} & \frac{33}{2} & -\frac{27}{5} & \frac{137}{180} & 0 & 0 \\
 \frac{137}{180} & -\frac{49}{60} & -\frac{17}{12} & \frac{47}{18} & -\frac{19}{12} & \frac{31}{60} & -\frac{13}{180} & 0 & 0 \\
 -\frac{13}{180} & \frac{19}{15} & -\frac{7}{3} & \frac{10}{9} & \frac{1}{12} & -\frac{1}{15} & \frac{1}{90} & 0 & 0 \\
 \frac{1}{90} & -\frac{3}{20} & \frac{3}{2} & -\frac{49}{18} & \frac{3}{2} & -\frac{3}{20} & \frac{1}{90} & 0 & 0 \\
 0 & \frac{1}{90} & -\frac{3}{20} & \frac{3}{2} & -\frac{49}{18} & \frac{3}{2} & -\frac{3}{20} & \frac{1}{90} & 0 \\
 0 & 0 & \frac{1}{90} & -\frac{3}{20} & \frac{3}{2} & -\frac{49}{18} & \frac{3}{2} & -\frac{3}{20} & \frac{1}{90} \\
 0 & 0 & \frac{1}{90} & -\frac{1}{15} & \frac{1}{12} & \frac{10}{9} & -\frac{7}{3} & \frac{19}{15} & -\frac{13}{180} \\
 0 & 0 & -\frac{13}{180} & \frac{31}{60} & -\frac{19}{12} & \frac{47}{18} & -\frac{17}{12} & -\frac{49}{60} & \frac{137}{180} \\
 0 & 0 & \frac{137}{180} & -\frac{27}{5} & \frac{33}{2} & -\frac{254}{9} & \frac{117}{4} & -\frac{87}{5} & \frac{203}{45} \\
\end{array}
\right),
\end{align}
\begin{align}
&D_9^1 = \left(
\begin{array}{ccccccccccc}
 -\frac{761}{280} & 8 & -14 & \frac{56}{3} & -\frac{35}{2} & \frac{56}{5} & -\frac{14}{3} & \frac{8}{7} & -\frac{1}{8} & 0 & 0 \\
 -\frac{1}{8} & -\frac{223}{140} & \frac{7}{2} & -\frac{7}{2} & \frac{35}{12} & -\frac{7}{4} & \frac{7}{10} & -\frac{1}{6} & \frac{1}{56} & 0 & 0 \\
 \frac{1}{56} & -\frac{2}{7} & -\frac{19}{20} & 2 & -\frac{5}{4} & \frac{2}{3} & -\frac{1}{4} & \frac{2}{35} & -\frac{1}{168} & 0 & 0 \\
 -\frac{1}{168} & \frac{1}{14} & -\frac{1}{2} & -\frac{9}{20} & \frac{5}{4} & -\frac{1}{2} & \frac{1}{6} & -\frac{1}{28} & \frac{1}{280} & 0 & 0 \\
 \frac{1}{280} & -\frac{4}{105} & \frac{1}{5} & -\frac{4}{5} & 0 & \frac{4}{5} & -\frac{1}{5} & \frac{4}{105} & -\frac{1}{280} & 0 & 0 \\
 0 & \frac{1}{280} & -\frac{4}{105} & \frac{1}{5} & -\frac{4}{5} & 0 & \frac{4}{5} & -\frac{1}{5} & \frac{4}{105} & -\frac{1}{280} & 0 \\
 0 & 0 & \frac{1}{280} & -\frac{4}{105} & \frac{1}{5} & -\frac{4}{5} & 0 & \frac{4}{5} & -\frac{1}{5} & \frac{4}{105} & -\frac{1}{280} \\
 0 & 0 & -\frac{1}{280} & \frac{1}{28} & -\frac{1}{6} & \frac{1}{2} & -\frac{5}{4} & \frac{9}{20} & \frac{1}{2} & -\frac{1}{14} & \frac{1}{168} \\
 0 & 0 & \frac{1}{168} & -\frac{2}{35} & \frac{1}{4} & -\frac{2}{3} & \frac{5}{4} & -2 & \frac{19}{20} & \frac{2}{7} & -\frac{1}{56} \\
 0 & 0 & -\frac{1}{56} & \frac{1}{6} & -\frac{7}{10} & \frac{7}{4} & -\frac{35}{12} & \frac{7}{2} & -\frac{7}{2} & \frac{223}{140} & \frac{1}{8} \\
 0 & 0 & \frac{1}{8} & -\frac{8}{7} & \frac{14}{3} & -\frac{56}{5} & \frac{35}{2} & -\frac{56}{3} & 14 & -8 & \frac{761}{280} \\
\end{array}
\right), \\
&D_9^2 = \frac{1}{h_\rho^2}\left(
\begin{array}{ccccccccccc}
 \frac{29531}{5040} & -\frac{962}{35} & \frac{621}{10} & -\frac{4006}{45} & \frac{691}{8} & -\frac{282}{5} & \frac{2143}{90} & -\frac{206}{35} & \frac{363}{560} & 0 & 0 \\
 \frac{363}{560} & \frac{8}{315} & -\frac{83}{20} & \frac{153}{20} & -\frac{529}{72} & \frac{47}{10} & -\frac{39}{20} & \frac{599}{1260} & -\frac{29}{560} & 0 & 0 \\
 -\frac{29}{560} & \frac{39}{35} & -\frac{331}{180} & \frac{1}{5} & \frac{9}{8} & -\frac{37}{45} & \frac{7}{20} & -\frac{3}{35} & \frac{47}{5040} & 0 & 0 \\
 \frac{47}{5040} & -\frac{19}{140} & \frac{29}{20} & -\frac{118}{45} & \frac{11}{8} & -\frac{1}{20} & -\frac{7}{180} & \frac{1}{70} & -\frac{1}{560} & 0 & 0 \\
 -\frac{1}{560} & \frac{8}{315} & -\frac{1}{5} & \frac{8}{5} & -\frac{205}{72} & \frac{8}{5} & -\frac{1}{5} & \frac{8}{315} & -\frac{1}{560} & 0 & 0 \\
 0 & -\frac{1}{560} & \frac{8}{315} & -\frac{1}{5} & \frac{8}{5} & -\frac{205}{72} & \frac{8}{5} & -\frac{1}{5} & \frac{8}{315} & -\frac{1}{560} & 0 \\
 0 & 0 & -\frac{1}{560} & \frac{8}{315} & -\frac{1}{5} & \frac{8}{5} & -\frac{205}{72} & \frac{8}{5} & -\frac{1}{5} & \frac{8}{315} & -\frac{1}{560} \\
 0 & 0 & -\frac{1}{560} & \frac{1}{70} & -\frac{7}{180} & -\frac{1}{20} & \frac{11}{8} & -\frac{118}{45} & \frac{29}{20} & -\frac{19}{140} & \frac{47}{5040} \\
 0 & 0 & \frac{47}{5040} & -\frac{3}{35} & \frac{7}{20} & -\frac{37}{45} & \frac{9}{8} & \frac{1}{5} & -\frac{331}{180} & \frac{39}{35} & -\frac{29}{560} \\
 0 & 0 & -\frac{29}{560} & \frac{599}{1260} & -\frac{39}{20} & \frac{47}{10} & -\frac{529}{72} & \frac{153}{20} & -\frac{83}{20} & \frac{8}{315} & \frac{363}{560} \\
 0 & 0 & \frac{363}{560} & -\frac{206}{35} & \frac{2143}{90} & -\frac{282}{5} & \frac{691}{8} & -\frac{4006}{45} & \frac{621}{10} & -\frac{962}{35} & \frac{29531}{5040} \\
\end{array}
\right).
\end{align}
\label{eq:finite_difference_operators_appendix}
\end{subequations}

%\bibliography{bibliography.bib}% Produces the bibliography via BibTeX.
%\bibliographystyle{jphysicsB}
\printbibliography
\end{document}